\documentclass[preprint,3p,12pt,authoryear]{elsarticle}
\usepackage{booktabs}
\usepackage{mathrsfs}
\usepackage{amssymb}
\usepackage{amsmath}
\usepackage{subfigure}
\usepackage{multirow}
\usepackage{threeparttable}
\usepackage{upgreek}

\begin{document}

\begin{frontmatter}

\title{In-orbit Calibration to the Point-Spread Function of \emph{Insight-HXMT}}

\author[a,b]{Yi Nang\corref{cor1}}
\ead{nangyi@ihep.ac.cn}
\author[a]{Jin-Yuan Liao\corref{cor1}}
\ead{liaojinyuan@ihep.ac.cn}
\author[a,b]{Na Sai}
\author[b,c]{Chen Wang}
\author[a]{Ju Guan}
\author[a]{Cheng-Kui Li}
\author[a,b]{Cheng-Cheng Guo}
\author[c]{Yuan Liu}
\author[a]{Jing Jin}
\author[a]{Xiao-Bo Li}

\author[a]{Shu Zhang}
\author[a,b,c]{Shuang-Nan Zhang}

\cortext[cor1]{Corresponding author}

\address[a]{Key Laboratory of Particle Astrophysics, Institute of High Energy Physics, Chinese Academy of Sciences, Beijing 100049, China}
\address[b]{University of Chinese Academy of Sciences, Beijing 100049, China}
\address[c]{Key Laboratory of Space Astronomy and Technology, National Astronomical Observatories, Chinese Academy of Sciences, Beijing 100012, China}

\begin{abstract}
We make the in-orbit calibration to the point-spread functions (PSFs) of the collimators of the \emph{Hard X-ray Modulation Telescope} with the scanning observation of the Crab. We construct the empirical adjustments to the theoretically calculated geometrical PSFs. The adjustments contain two parts: a rotating matrix to adjust the directional deviation of the collimators and a paraboloidal function to correct the inhomogeneity of the real PSFs. The parameters of the adjusting matrices and paraboloidal functions are determined by fitting the scanning data with lower scanning speed and smaller intervals during the calibration observations. After the PSF calibration, the systematic errors in source localization in the Galactic plane scanning survey are $0^{\circ}.010 $, $0^{\circ}.015$, $0^{\circ}.113$ for the Low-Energy Telescope (LE), the Medium-Energy telescope (ME) and the High-Energy telescope (HE), respectively; meanwhile, the systematic errors in source flux estimation are  1.8\%, 1.6\%, 2.7\% for LE, ME and HE, respectively.
\end{abstract}


\end{frontmatter}

\section{Introduction}

\begin{figure}
    \centering
        \includegraphics[width=6.5cm]{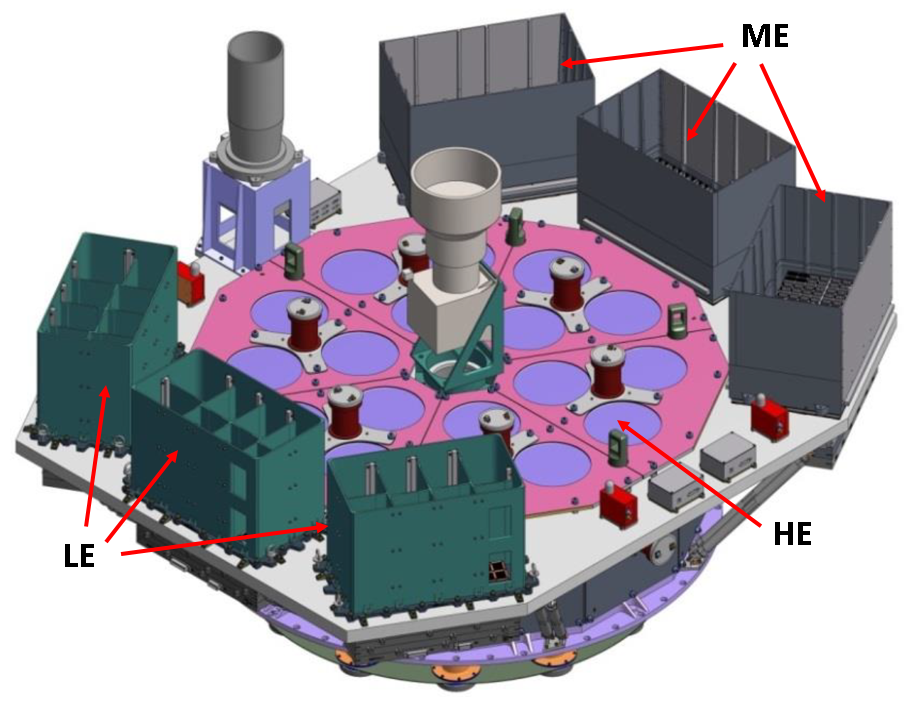}
        \caption{The main structure of \emph{Insight-HXMT}.}
        \label{fig_allstru}
    \end{figure}

\begin{figure}
    \centering
    \subfigure[LE]{
        \includegraphics[width=5 cm]{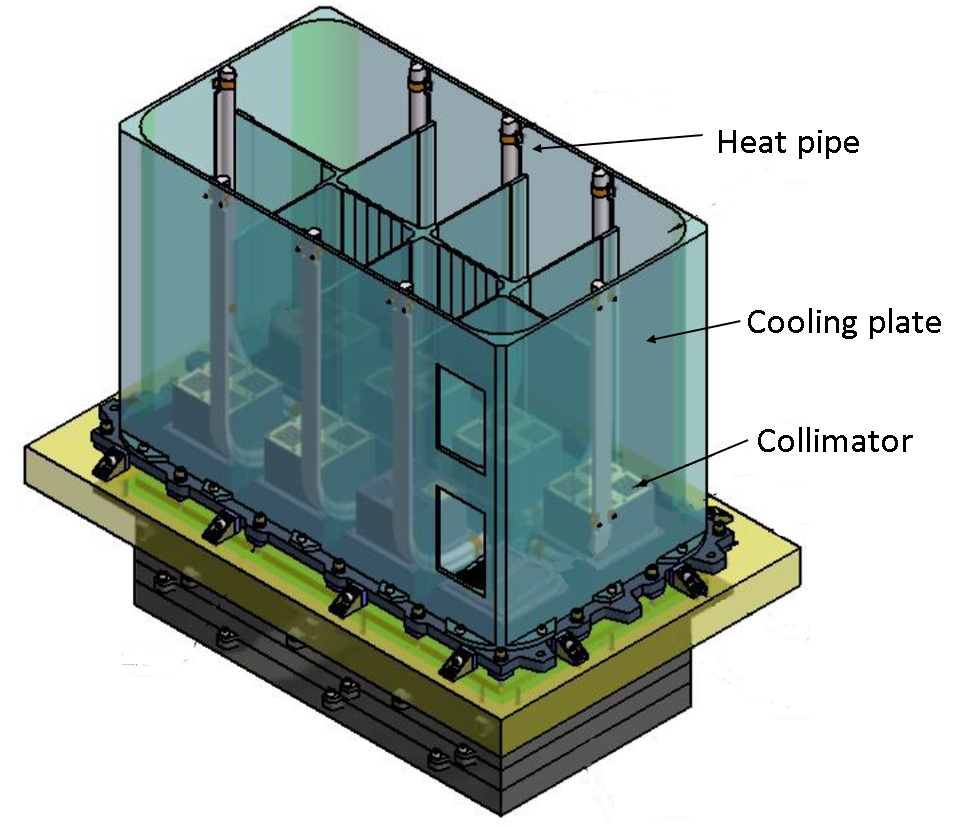}
    }
    \subfigure[ME]{
        \includegraphics[width=5 cm]{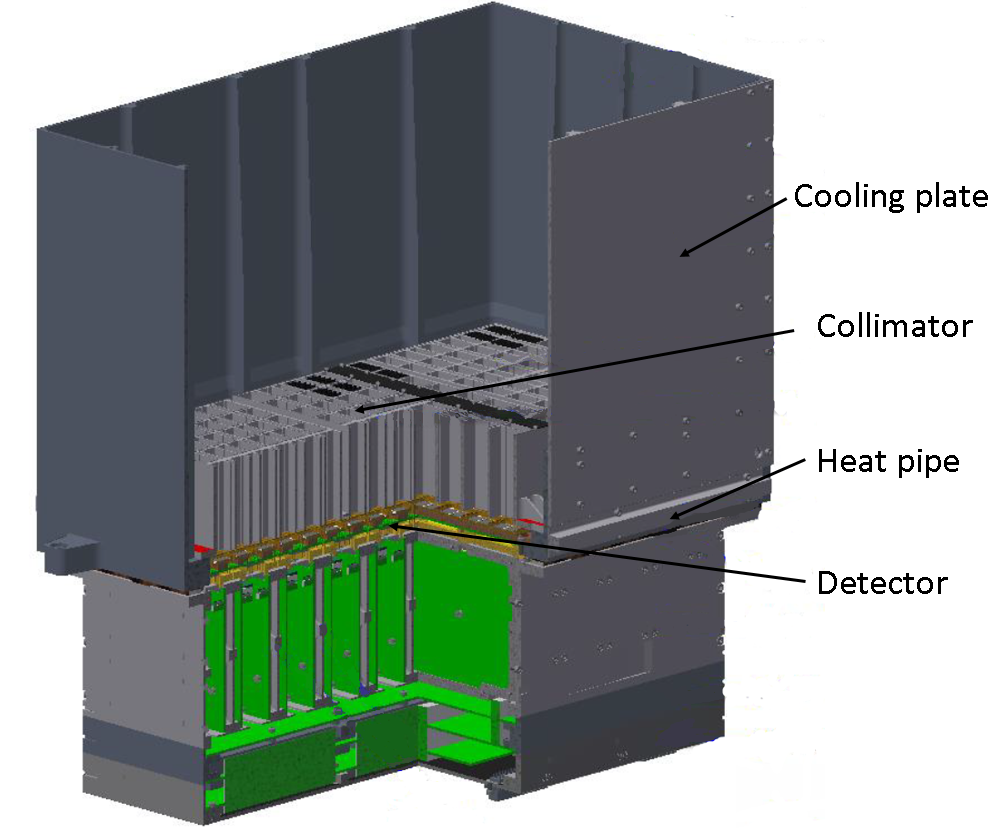}
    }
    \subfigure[HE]{
        \includegraphics[width=5 cm]{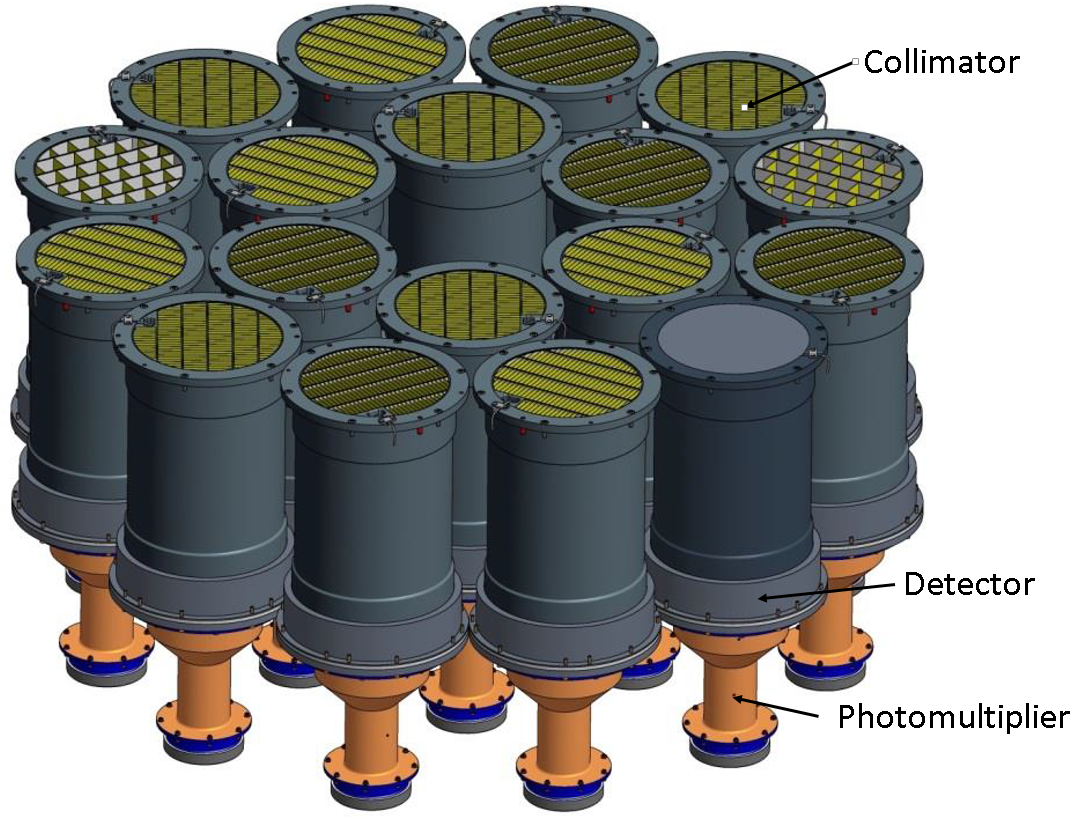}
    }
    \caption{The structures of LE, ME and HE modules.}
    \label{fig_box}
\end{figure}

The \emph{Hard X-ray Modulation Telescope} (\emph{Insight-HXMT}) is China's first X-ray telescope based on the Direct Demodulation Method \citep{Li1993,Li1994} and was launched on June 15th, 2017. \emph{Insight-HXMT} carries three main instruments with different energy ranges: Low Energy X--ray Telescope (LE), Medium Energy X--ray Telescope (ME) and High Energy X-ray Telescope (HE) \citep{Zsn2019}. The main structure of \emph{Insight-HXMT} and the three telescope modules are shown in Figures \ref{fig_allstru} and \ref{fig_box}. LE is composed of 96 swept charge devices (SCD) that are sensitive in 0.7--13~keV with a total geometrical area of 384 cm$^{2}$. ME consists of 1728 pixels of Si-PIN detectors covering the energy range of 5--40~keV with a total geometrical area of 952 cm$^{2}$. HE consists of 18 NaI/CsI detectors covering the energy range of 20--250~keV with a total geometrical area of 5096 cm$^{2}$. The details of LE, ME and HE have been described by \citet{Cy2019}, \citet{Cxl2019} and \citet{Lcz2019}.

\begin{figure}
    \centering
    \subfigure{
        \includegraphics[width=6.5cm]{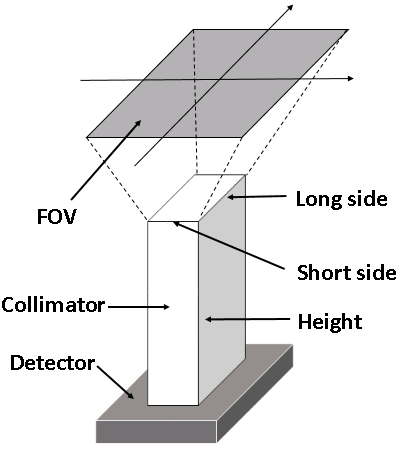}
    }
    \subfigure{
        \includegraphics[width=6.5cm]{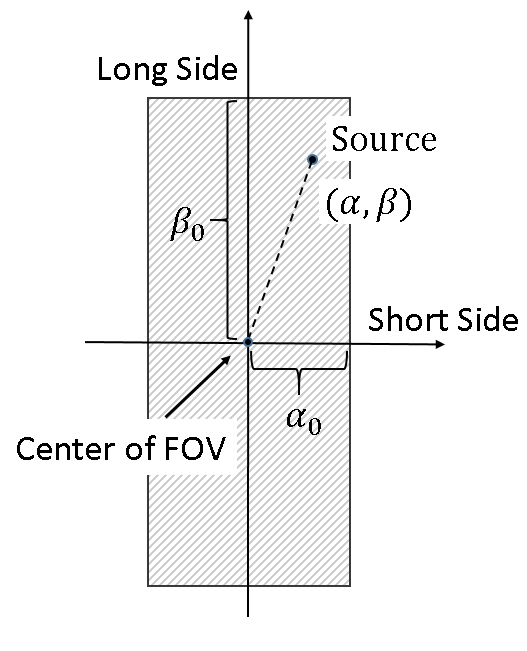}
    }
    \caption{The Left panel is the illustration of the rectangular grid in the collimators of \emph{Insight-HXMT}. The right panel shows the coordinates of the FOV, where $(\alpha,\beta)$ is the position of a source in the FOV. $\alpha_{0}$ and $\beta_{0}$ are the sizes of the FOV in the short and long side.}
    \label{fig_fov}
    \end{figure}

All the three instruments are slat-collimated type of telescopes. Rectangular metallic grid collimators are placed on the detectors of LE, ME and HE to shield the photons outside the field of views (FOVs). The materials of the collimator walls for LE, ME and HE are 0.12~mm aluminum, 0.07~mm tantalum and 0.2~mm tantalum, respectively. Figure \ref{fig_fov} shows the geometrical structure of individual grid in the collimator and the coordinates of the FOV.
Each telescope is mostly composed of detectors with small FOV and supplemented with large FOV detectors. The detailed attributes of \emph{Insight-HXMT} are shown in Table \ref{tab_instr}.

\begin{table}[tp]
    \centering
    \caption{Attributes of \emph{Insight-HXMT}}
    \label{tab_instr}
    \begin{threeparttable}
    \begin{tabular}{ccccc}
        \toprule
        \multicolumn{2}{c}{Telescope} & LE & ME & HE \\
        \midrule
        \multicolumn{2}{c}{Type} & SCD & Si-PIN & NaI(Tl)/CsI(Na) \\
        \multicolumn{2}{c}{Geometrical Area $(\textrm{cm}^{2})$} & 384 & 952 & 5096 \\
        \multicolumn{2}{c}{Collimator Wall Material} & Aluminum & Tantalum & Tantalum \\
        \multicolumn{2}{c}{Collimator Wall Thickness} & 0.12~mm & 0.07~mm & 0.14~mm \\
        \multicolumn{2}{c}{Energy Range (keV)\tnote{a}} & $0.7-13$ & $5-40$ & $20-250$ \\
        \multicolumn{2}{c}{Energy Range (keV)\tnote{b}} & $1-6$ & $7-40$ & $25-100$ \\
        \multirow{2}{*}{\qquad  FOV (FWHM)} &Small& $1^{\circ}.6\times 6^{\circ}$ & $ 1^{\circ} \times 4^{\circ}$ & $1^{\circ}.1\times 5^{\circ}.7$ \\
         &Large & $4^{\circ} \times 6^{\circ}$ & $4^{\circ} \times 4^{\circ}$ & $5^{\circ}.7 \times 5^{\circ}.7$\\
        \bottomrule
        \end{tabular}
\begin{tablenotes}
\item[a] The whole effective energy ranges of \emph{Insight-HXMT}.
\item[b] The energy ranges used in the Galactic plane scanning survey in post-processing the data on ground.
\end{tablenotes}
\end{threeparttable}

\end{table}

The Galactic plane scanning survey is one of the core science programs of \emph{Insight-HXMT}. The main task is to find new transient sources and to monitor known variable sources in the Galactic plane. In a scanning observation, the FOVs of the telescope sweep across sky, and there will be a count rate peak left on the light curve when a source crosses each FOV. The shape of the count rate peak depends on the point-spread function (PSF) of the collimator. Therefore, the position and flux of the source can be inferred from the fitting of the count rate peak with the PSF and attitude of the telescope. However, there are various factors like vibration and thermal deformation that would affect PSF after the \emph{Insight-HXMT} launch. These effects result in large systematic errors in the source localization and flux estimation. Consequently, an in-orbit PSF calibration is essential to obtain accurate and reliable results from the scanning observations of \emph{Insight-HXMT}. Almost all X-ray and gamma-ray telescopes performed the alignment and PSF calibrations after launch, such as \emph{HEAO-1} \citep{Roy1977}, \emph{Fermi} \citep{Ackermann2012}, \emph{Nustar} \citep{Madsen2015} and MAXI \citep{Hiroi2013}.

In this paper, we use a series of scanning observations of the Crab nebula and pulsar (hereafter the ``Crab'') as calibrating data to analyse the PSF transformation, and we correct the transformation with two groups of parameters: three Euler angles to correct the rotational distortion of the collimators and a paraboloidal function with four parameters to correct other inhomogeneous transformation of the PSFs. Finally, we use the regular scanning observations of the Crab to test the calibrated PSFs. This paper is organized as follows. The observational information and data reduction are presented in Section 2. The model adjustment is described in Section 3. In Section 4, we present the result of the calibration. Finally, a summary and a discussion are given in Section 5.

\section{Observations and Data Reduction}
All the three telescopes are mostly composed of detectors with small FOV as described in Section 1. Compared to the large FOV detectors, the small FOV detectors can give more accurate source position thanks to the relatively narrow FOVs. In addition, the background of the large FOV detectors is more complicated than that of the small FOV detectors. Thus, the small FOV detectors of the three telescopes are used to do the Galactic plane scanning survey currently. In this paper, we focus on the PSF calibration of the small FOV detectors.

\begin{figure}
    \centering
        \includegraphics[width=6.5cm]{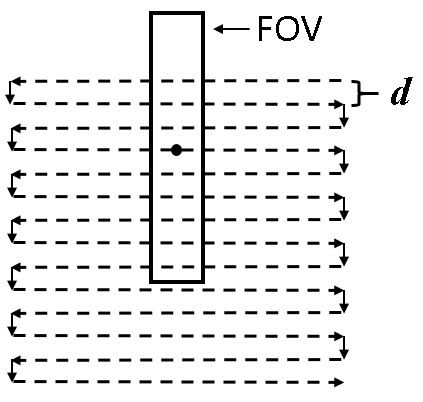}
        \caption{Schematic diagram of a normal scanning observation. The dashed line is the trajectory of the FOV center. The area of the observation is about $7^{\circ}\times7^{\circ}$. $d$ is the interval between the scanning tracks. In normal scanning observations, $d=0^{\circ}$.4.}
        \label{fig_scan}
    \end{figure}

The Crab is used as a calibrating source due to its high and stable flux in the three energy ranges of \emph{Insight-HXMT}. Although the Crab's extension is about $2^\prime$ at $\sim 1$~keV, it still can be considered as a point source by \emph{Insight-HXMT} which has broad PSFs larger than $1^{\circ}$. Two types of scanning data of the Crab are used to do PSF calibration and test. In regular scanning observations, the standard scanning speed $v$ is $0^{\circ}.06~{\rm s}^{-1}$, and the scanning interval $d$ is $0^\circ.4$. We use the special scanning observations of the Crab with lower speed and smaller interval ($v=0^{\circ}.03~{\rm s}^{-1}$ \& $ d=0^\circ.1$ ) performed on Oct. 2017 and Oct. 2018 to calibrate the PSFs. The regular scanning observations of the Crab are used to test the performance of the PSF calibration. Figure \ref{fig_scan} shows the schematic diagram of scanning observations.

The energy ranges used in the Galactic plane scanning survey are 1--6~keV, 7--40~keV and 25--100~keV for LE, ME and HE, respectively. The data reduction is performed by the \emph{Insight-HXMT} data analysis software package HXMTDAS v2.0 that mainly contains the following steps:
\begin{itemize}
    \item PI transformation to generate calibrated events with the HXMTDAS tasks of \textbf{hepical}, \textbf{mepical}, and \textbf{lepical} tasks for HE, ME, and LE instruments, respectively.
    \item Event reconstruction for LE with \textbf{lerecon} task and event classify for ME with \textbf{megrade} task.
    \item Good Time Interval (GTI) selection considering geomagnetic cutoff rigidity, elevation angles and the south Atlantic Anomaly with \textbf{hegtigen}, \textbf{megtigen}, and \textbf{legtigen} tasks. 
    \item Event screening to select calibrated events according to GTIs with \textbf{hescreen}, \textbf{mescreen}, and \textbf{lescreen} tasks.
    \item Generating light curve from screening files with \textbf{helcgen}, \textbf{melcgen} and \textbf{lelcgen} tasks.
\end{itemize}
   
The background of HE and ME are remarkably modulated by the geomagnetic field \citep{Li2009,Xie2015}. We use polynomial fit to estimate the backgrounds of HE and ME. For LE, although the background is generally at low level, it becomes volatile at high level due to particle events. The Statistics-sensitive Nonlinear Iterative Peak (SNIP) method \citep{Ryan1988, Morh1997}, which is a background approximation method used in analyzing X-ray spectral, is used to estimate the background of LE.

\section{Data Analysis and PSF Adjustment}
    
\begin{figure}
    \centering
    \subfigure[LE]{
        \includegraphics[width=5cm,height=6cm]{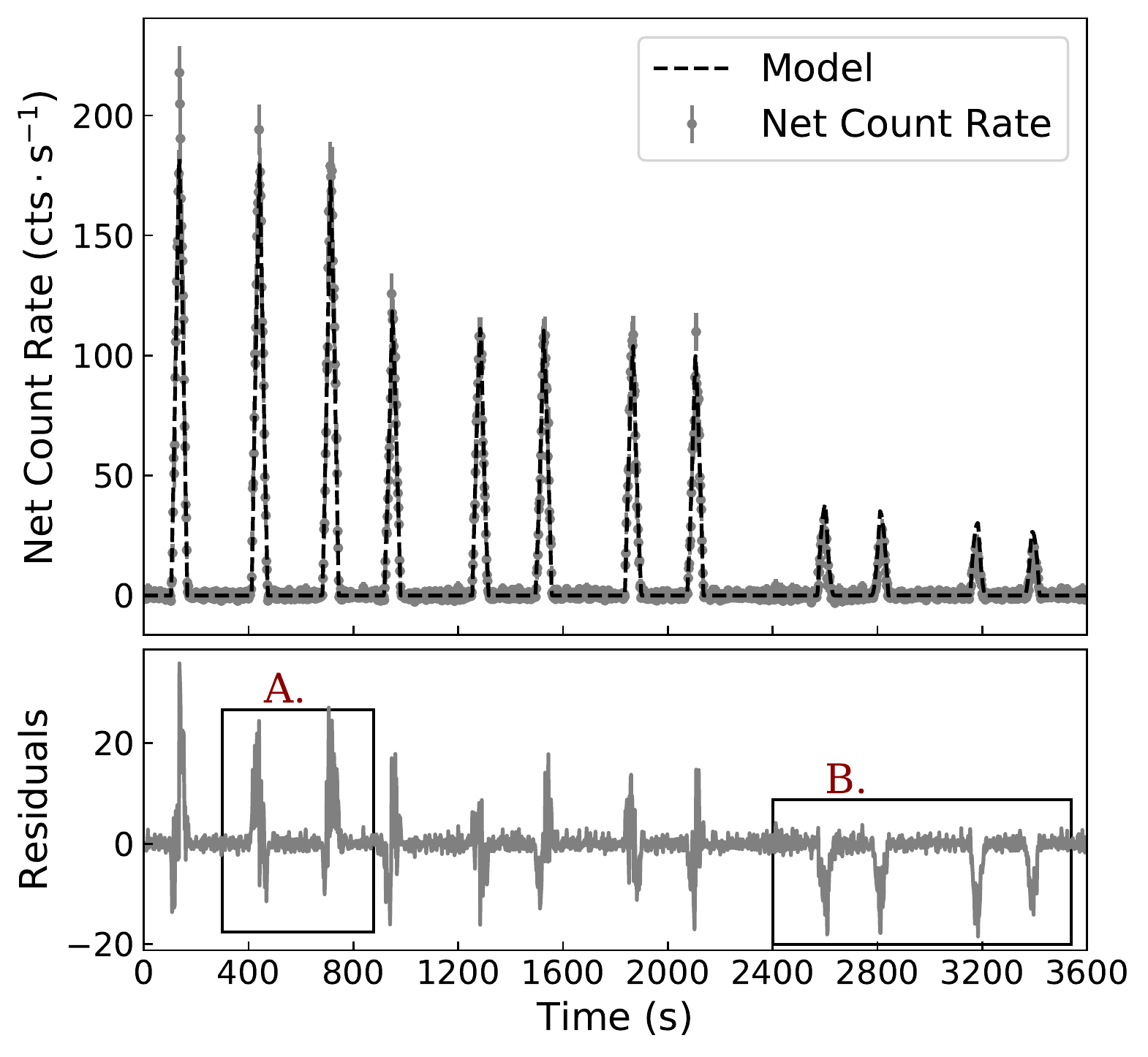}
    }
    \subfigure[ME]{
        \includegraphics[width=5cm,height=6cm]{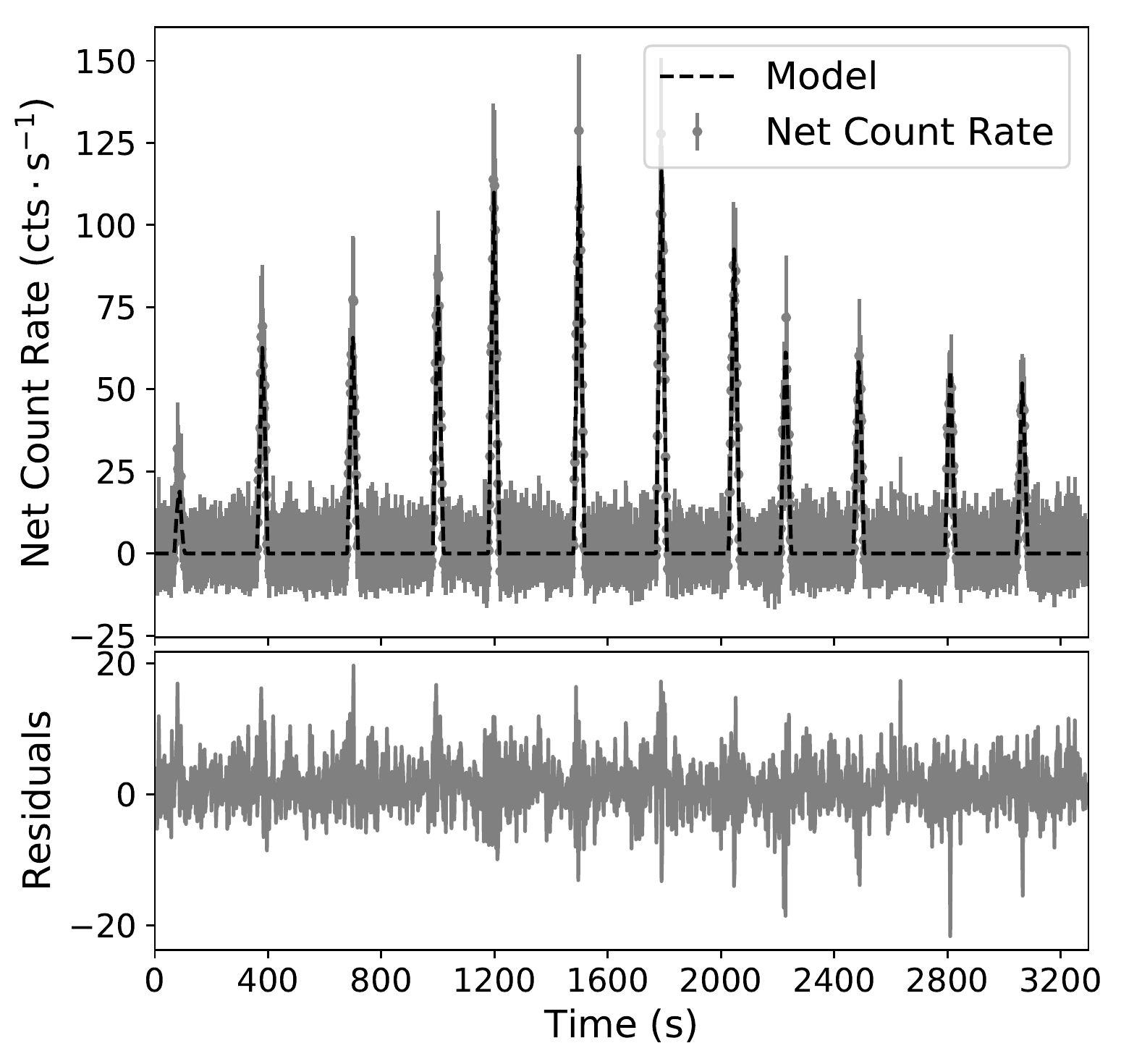}
    }
    \subfigure[HE]{
        \includegraphics[width=5cm,height=6cm]{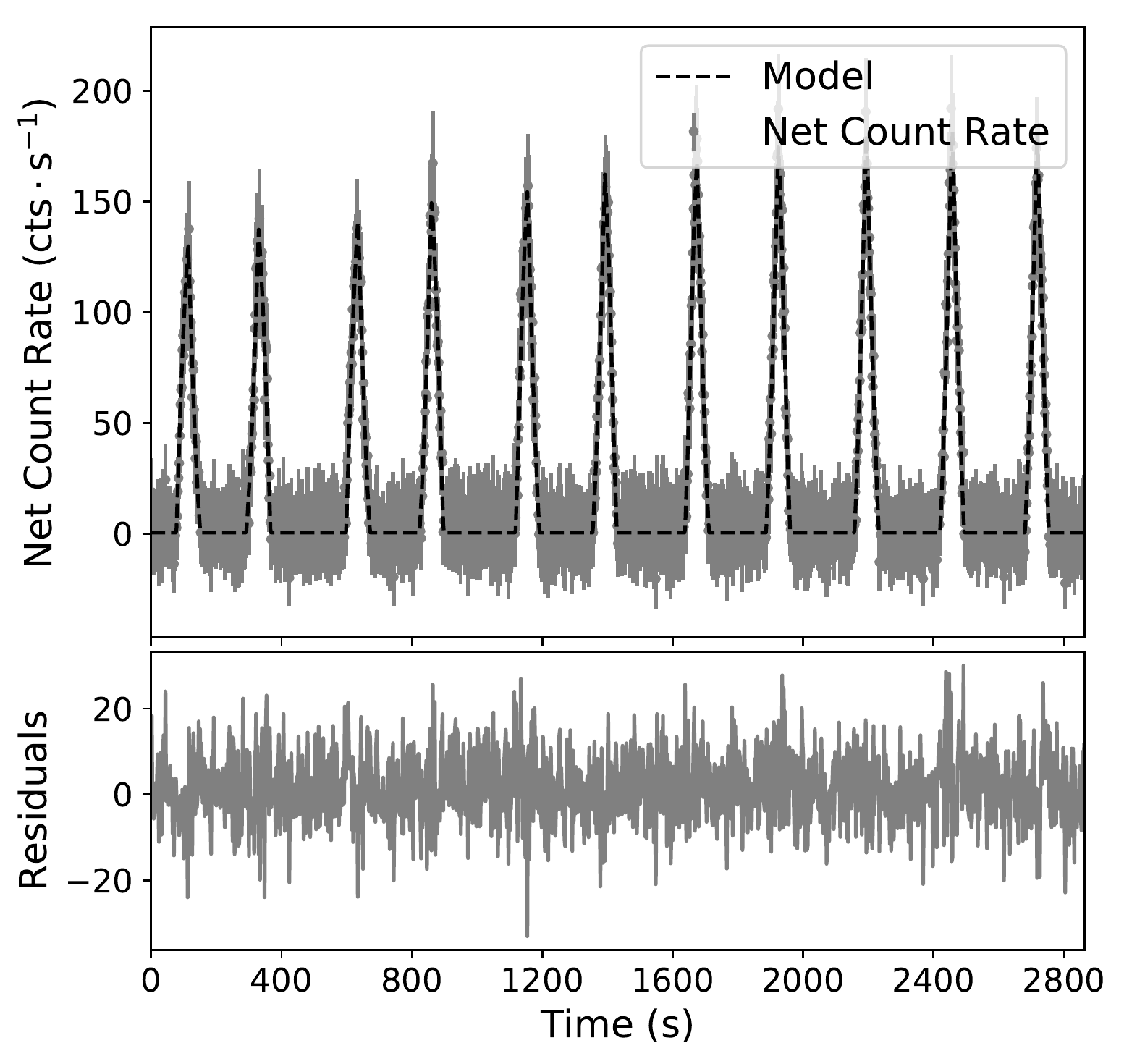}
    }
    \caption{Scanning data fitting with the geometrical PSF. The top panel is the PSF fitting to the net light curve of a scanning observation around the Crab. The bottom panel shows the fitting residuals. Marked structures exist in the residuals of LE data that mean the data is not well-fitted. The fittings of ME and HE is better than LE, but still can be improved.}
    \label{fig_example}
\end{figure}

The collimators of the detectors define the FOVs and PSFs of the telescopes. According to the calibrating experiments on ground, PSFs in different energies within the ranges we selected are generally similar. The PSFs inferred from the geometry of collimators can be described as
\begin{equation}
    \mathcal{P}(\alpha,\beta) = C \times \frac{(1-\frac{abs(\tan(\alpha))}{\tan(\alpha _{0})}) \times (1-\frac{abs(\tan(\beta))}{\tan(\beta _{0})})}{\sqrt{\tan ^{2}(\alpha) + \tan ^{2}(\beta)+1}},
    \label{eqno1}
\end{equation}
where $\alpha$ and $\beta$ are the source's position relative to the long side and short side of a telescope, respectively, $\alpha _{0}$ and $\beta _{0}$ are the sizes of the long side and short side of the FOV (FWHM) which are listed in Tab. \ref{tab_instr} and illustrated in Fig. \ref{fig_fov}, $C$ is the normalized flux of the source. This equation simply describes the ratio of the area that can receive photons to the total detector area.

The flux and position of a source in a scanned region are estimated by fitting the net light curve with its PSF. The position (right ascension and declination, R.A.and Dec.) and normalized flux are parameters in the fitting. Fig. \ref{fig_example} shows the fittings for HE, ME and LE light curve with geometrical PSF described in Equ. \ref{eqno1}. The scanning data can not be well-fitted with the geometrical PSFs directly, especially for LE data. When the R.A. and Dec. of the Crab are fixed to its real position, marked structures exist in the fitting residuals. It can be a result of thermal deformation or vibration that make real PSFs deviate from geometrical settings after the launch. HE and ME have similar problems but are not as serious as LE. According to the fitting result of the calibrating data, we correct the deviations with two part corrections: rotating correction and inhomogeneity correction.

\begin{figure}
    \centering
    \subfigure[LE]{
        \includegraphics[width=5cm,height=6cm]{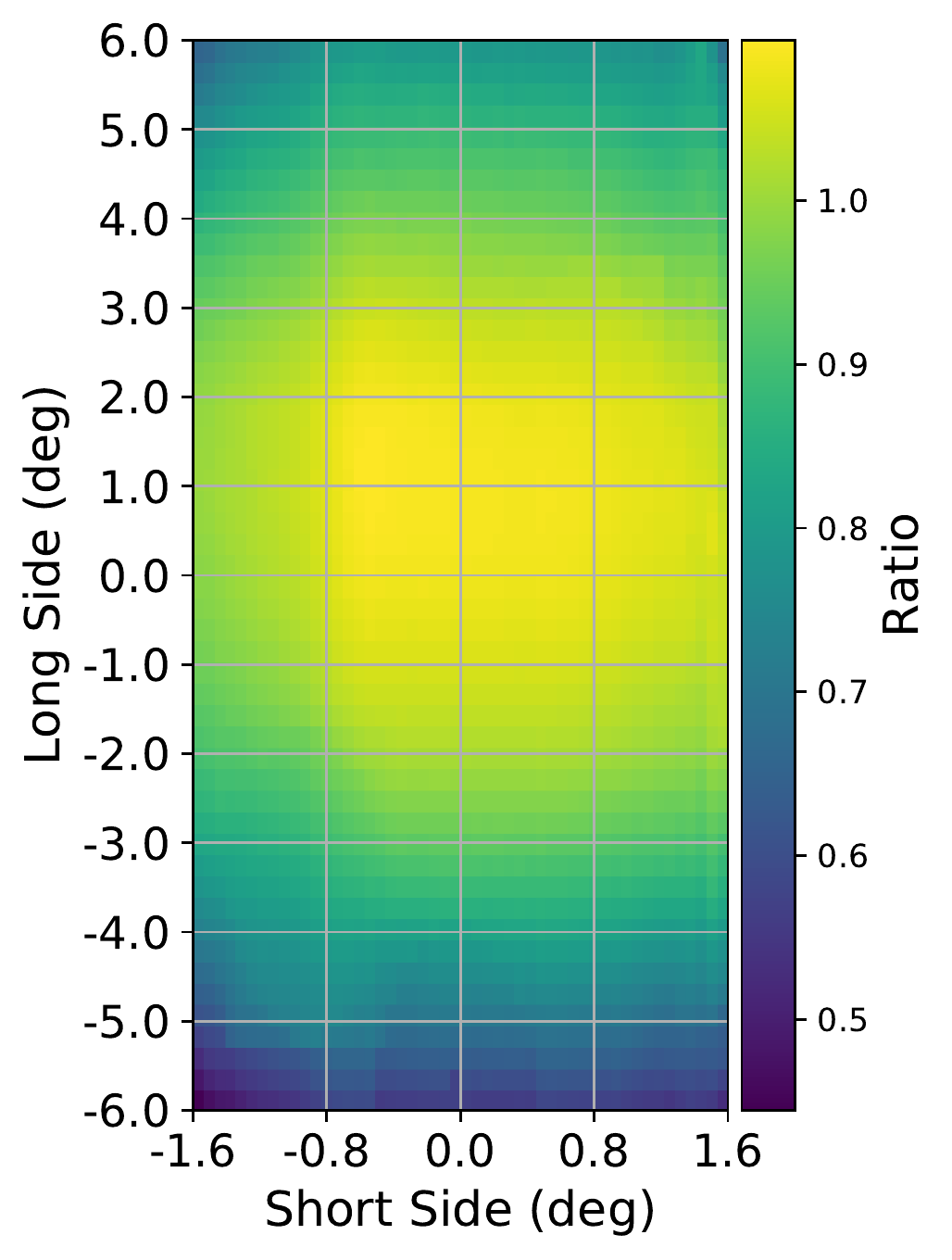}
    }
    \subfigure[ME]{
        \includegraphics[width=5cm,height=6cm]{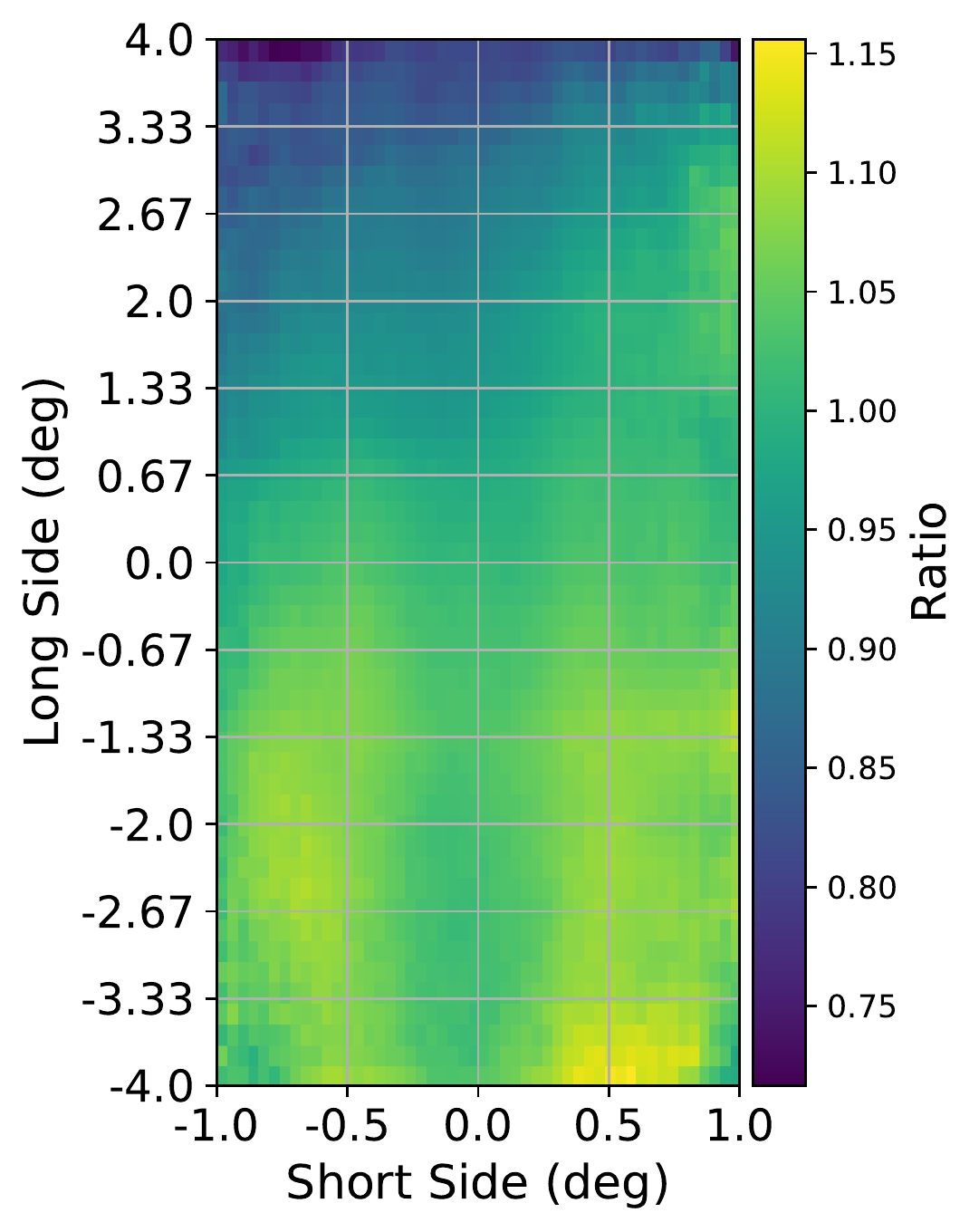}
    }
    \subfigure[HE]{
        \includegraphics[width=5cm,height=6cm]{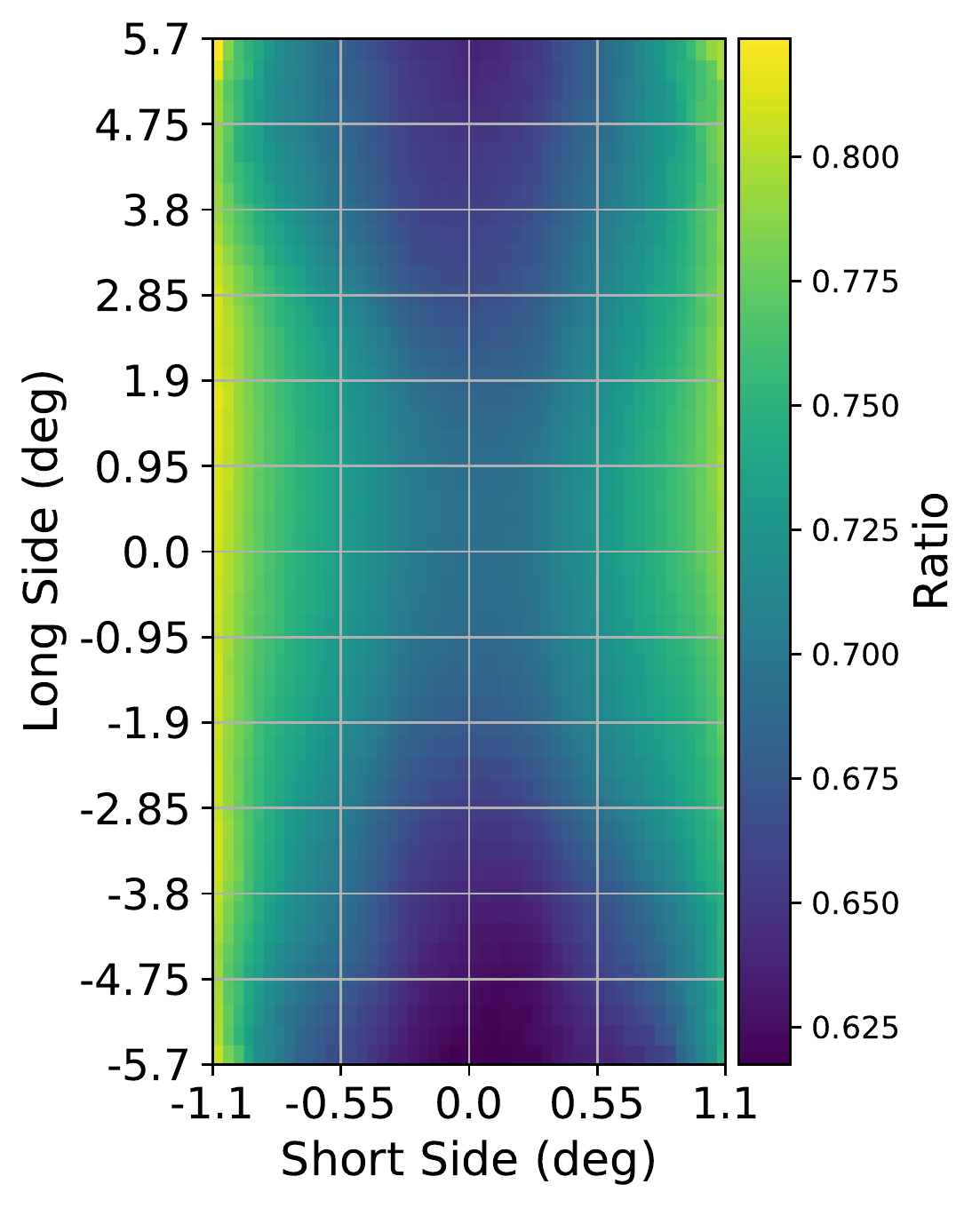}
    }
    \caption{Ratios between observed count rate and fitting result with the geometrical PSF in the different orientations of FOVs. Ratio greater than unity indicates that the observed count rate is higher than fitting value using geometrical PSF. The ratios are interpolated and smoothed to make up for the deficiency of calibrating data in several orientations.}
    \label{fig_ratio}
\end{figure}

\begin{enumerate}[(1)]
    \item Rotating correction. In Fig. \ref{fig_example} (a), the predicted peaks of the models always deviate from the peaks in the data. A feature can be seen in Box A that two adjacent peaks have mirror symmetrical fitting residuals. If the collimators turn aside from the setting directions, residuals will have this types of features. The original source position in FOV can be calculated by
\begin{gather}
    \alpha = \arctan\frac{x}{z},\quad \beta = \arctan\frac{y}{z}
\end{gather}
where $ x, y, z$ are the source orientation in the Cartesian coordinate of the telescope. A rotating matrix $M(\psi,\theta,\phi)$ with three Euler angles for rotating correction is invoked to correct the source position in FOV, and the corrected source position $(\alpha',\beta')$ can be calculated by
\begin{gather}
    \alpha' = \arctan\frac{x'}{z'},\quad \beta' = \arctan\frac{y'}{z'}\\
    \begin{bmatrix}x'\\y'\\z'\\ \end{bmatrix} = M(\psi,\theta,\phi) \cdot \begin{bmatrix}x\\y\\z\end{bmatrix}
\end{gather}
where $ \begin{bmatrix}x,&y,&z\end{bmatrix}$ is the source orientation in the Cartesian coordinate of the telescope and
\begin{gather}
     M(\psi,\theta,\phi)= 
\begin{bmatrix}
\cos\psi\cos\theta, & \sin\psi\cos\theta, &-\sin\theta \\
\cos\psi\sin\theta\sin\phi - \sin\psi\cos\phi, & \sin\psi\sin\theta\sin\phi+\cos\psi\cos\phi, & \cos\theta\sin\psi \\
\cos\psi\sin\theta\cos\phi + \sin\psi\sin\phi, & \sin\psi\sin\theta\cos\phi-\cos\psi\sin\phi, & \cos\theta\cos\psi \\
\end{bmatrix}.
\label{eq_rot}
\end{gather}

\item Inhomogeneity correction. In Box A and Box B of Fig. \ref{fig_example} (a), the residuals in different counts rate deviate in different ways. The observed count rate is higher than its fitting result when the count rate is high for Box A. On the contrary, the observed count rate is lower than the fitting result when the count rate is low for Box B. This feature also can be seen in Fig. \ref{fig_ratio} that shows the ratios between the values of data and fitting with the geometrical PSF in the different directions of FOVs. For LE, the ratio is higher in the center than at the edge. For HE, the ratio is higher in the left and right than the top and bottom. We choose a paraboloidal function with four parameters $(a,b,c,d)$ to describe this inhomogeneous feature. 
\end{enumerate}
\begin{gather}
    F(\alpha',\beta') = {a} \times \alpha'^{2} \times \beta'^{2} + {b} \times \alpha'^{2}+ {c} \times \beta'^{2}+{d},
   \label{eq_inhom}
\end{gather}

Finally, The corrected PSF $\mathcal{P'(\alpha,\beta) }$ can be described as follows

\begin{gather}
    \mathcal{P'}(\alpha,\beta) = \mathcal{P}(\alpha',\beta') \times F(\alpha',\beta').
    \label{eq_psf_c}
\end{gather}
where $\mathcal{P(\alpha',\beta')}$ is the geometrical PSF (Equation \ref{eqno1}) with the corrected position $(\alpha',\beta')$.

\section{Result}
\begin{figure}
    \centering
    \subfigure[LE]{
        \includegraphics[width=5 cm]{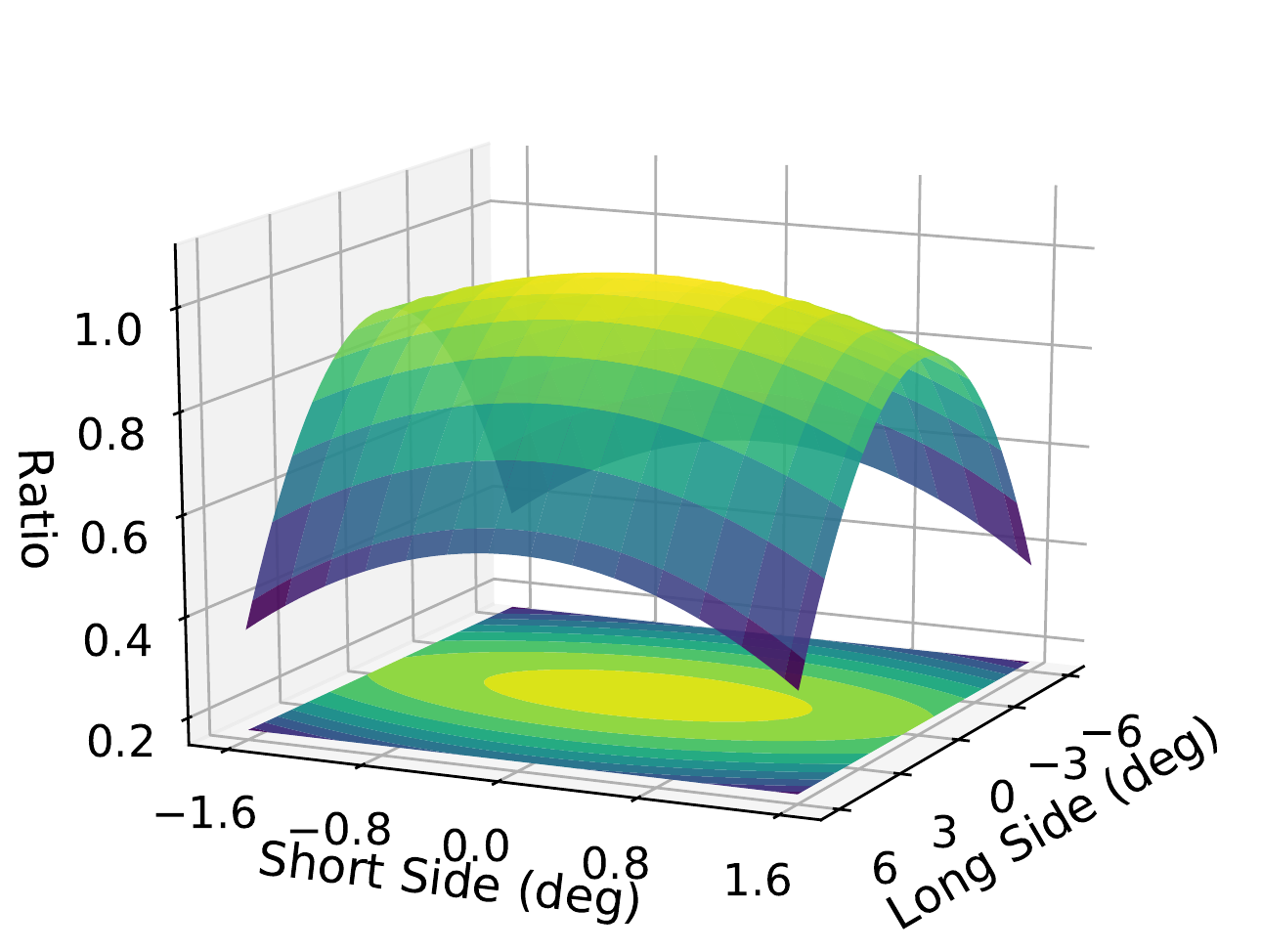}
    }
    \subfigure[ME]{
        \includegraphics[width=5 cm]{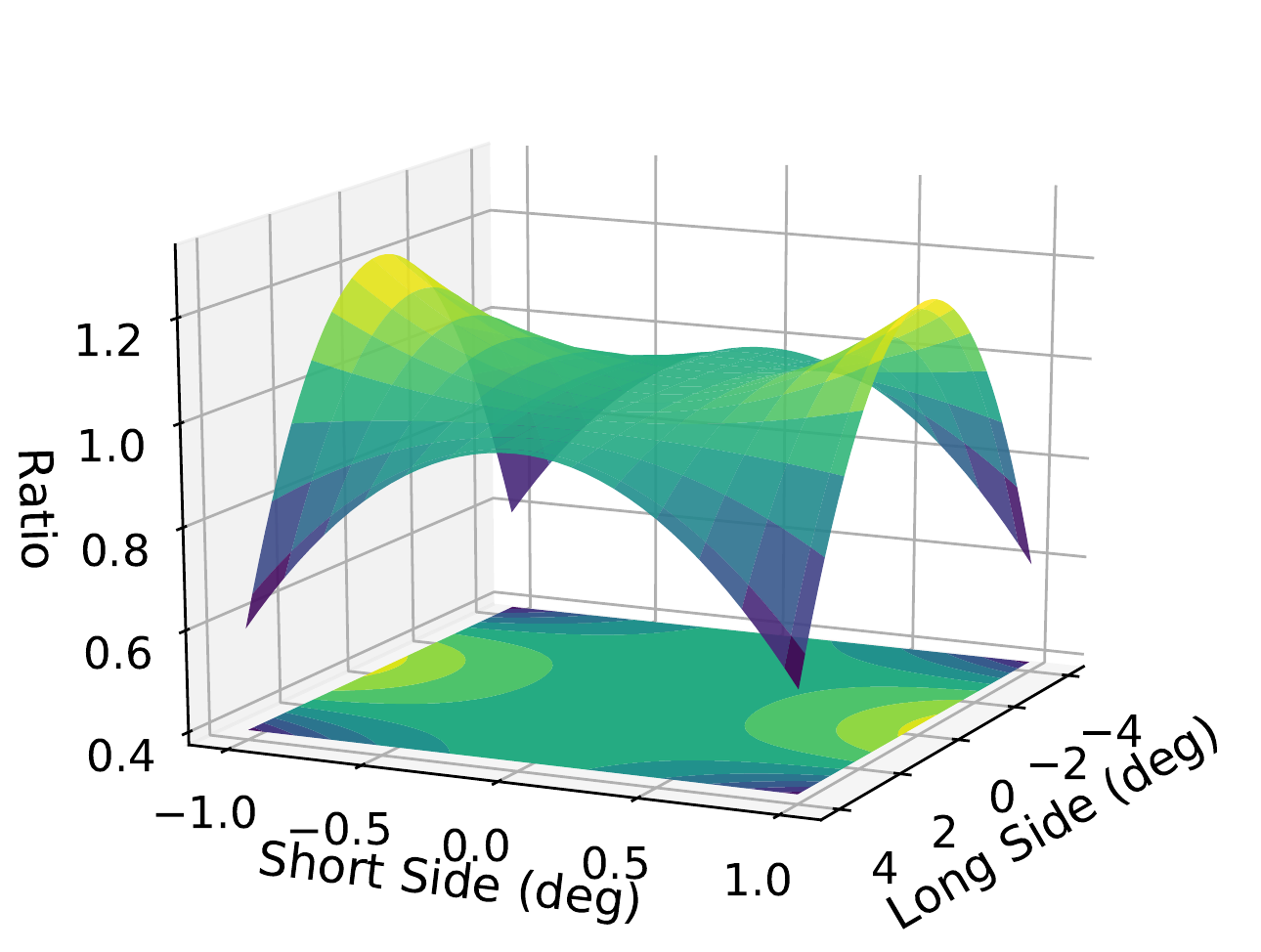}
    }
    \subfigure[HE]{
        \includegraphics[width=5 cm]{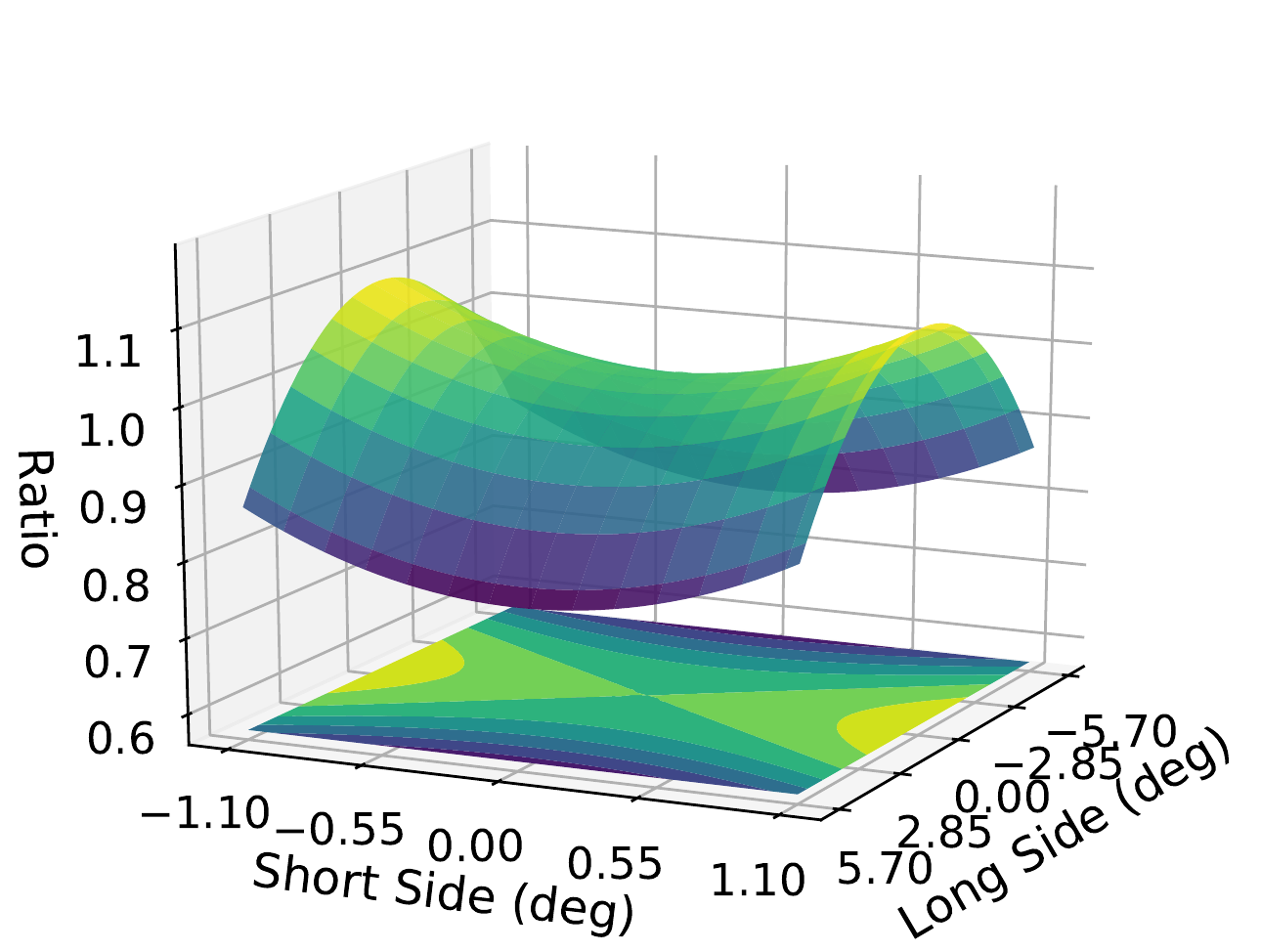}
    }
    \caption{Profiles of the paraboloidal function of LE, ME and HE.}
    \label{fig_fuc}
\end{figure}

\begin{figure}
    \centering
    \subfigure[LE]{
        \includegraphics[width=5cm,height=6cm]{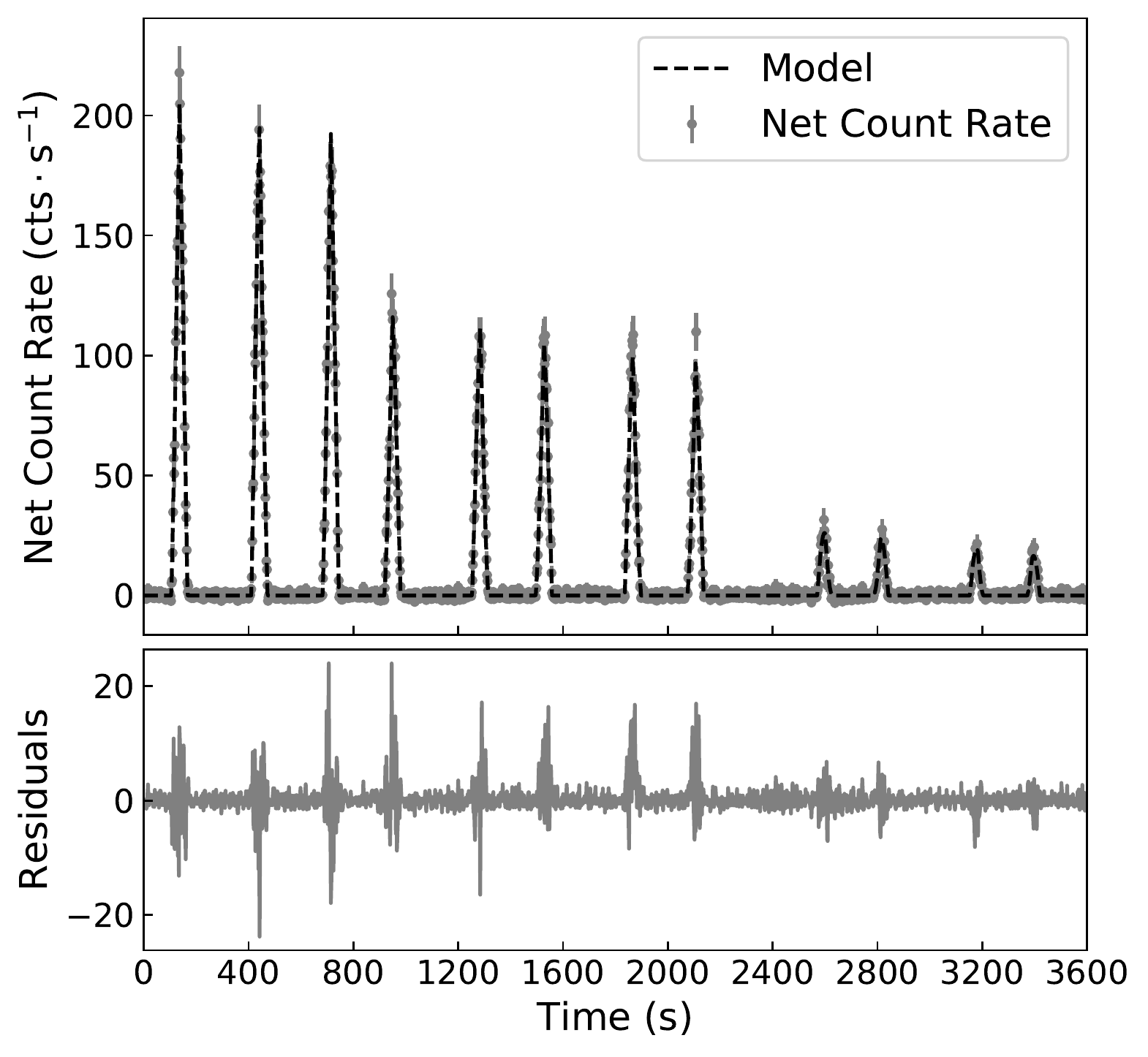}
    }
    \subfigure[ME]{
        \includegraphics[width=5cm,height=6cm]{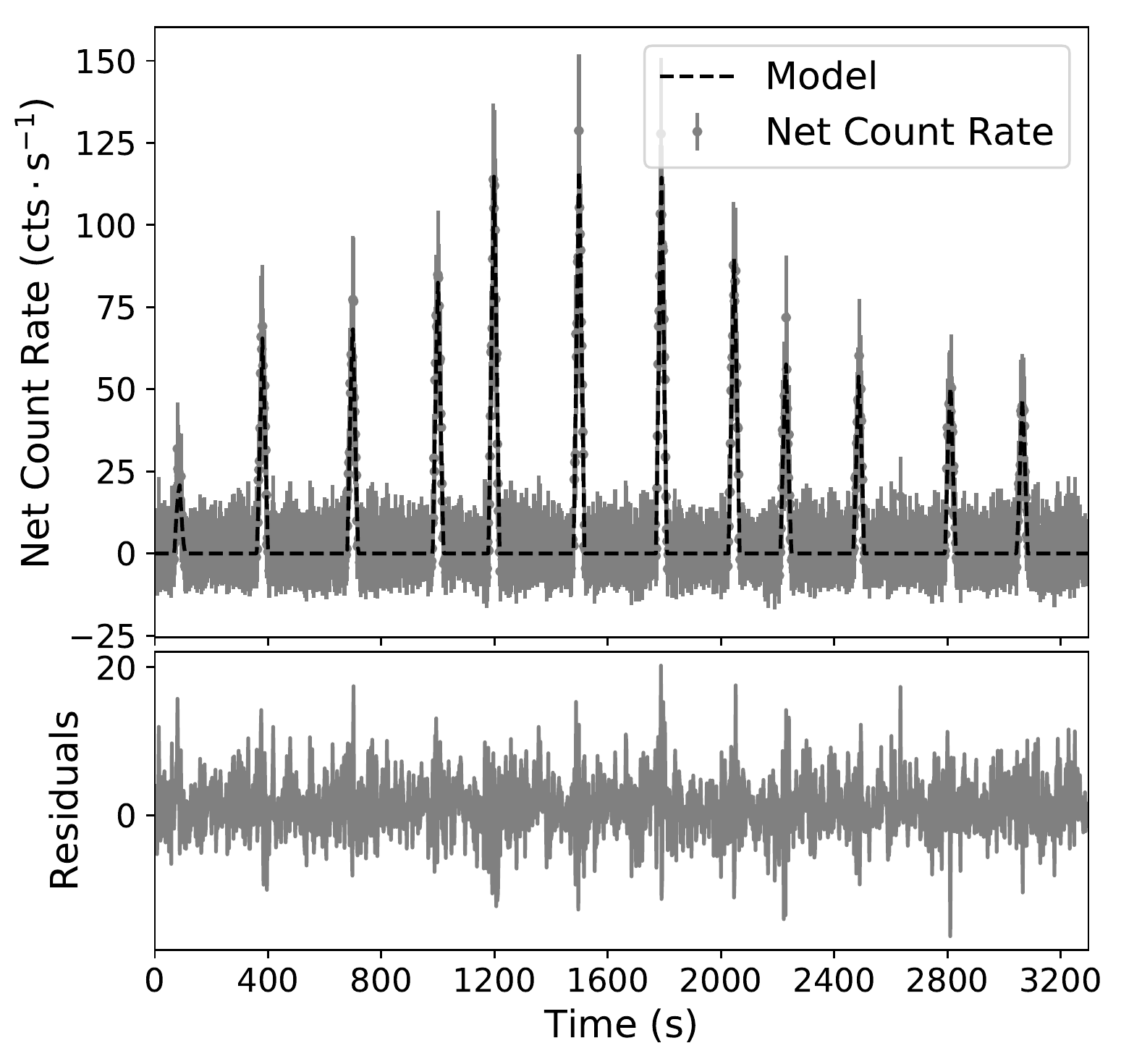}
    }
    \subfigure[HE]{
        \includegraphics[width=5cm,height=6cm]{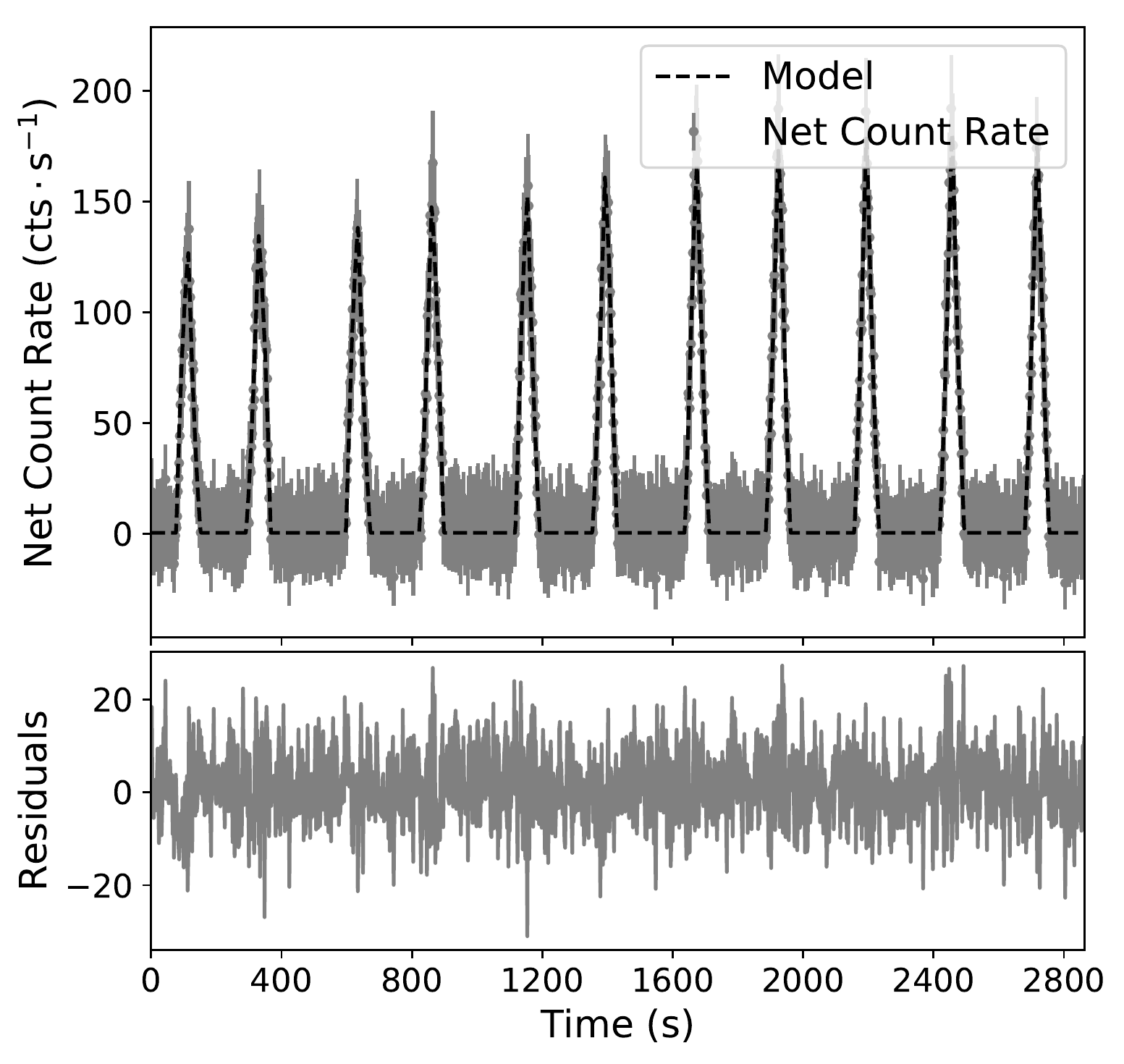}
    }
    \caption{The same as Figure \ref{fig_example}, but with the calibrated PSF. The fitting is better than that in Figure \ref{fig_example}.}
    \label{fig_new}
\end{figure}

\begin{figure}
    \centering
    \subfigure[LE: only with rotating correction]{
        \includegraphics[width=6 cm]{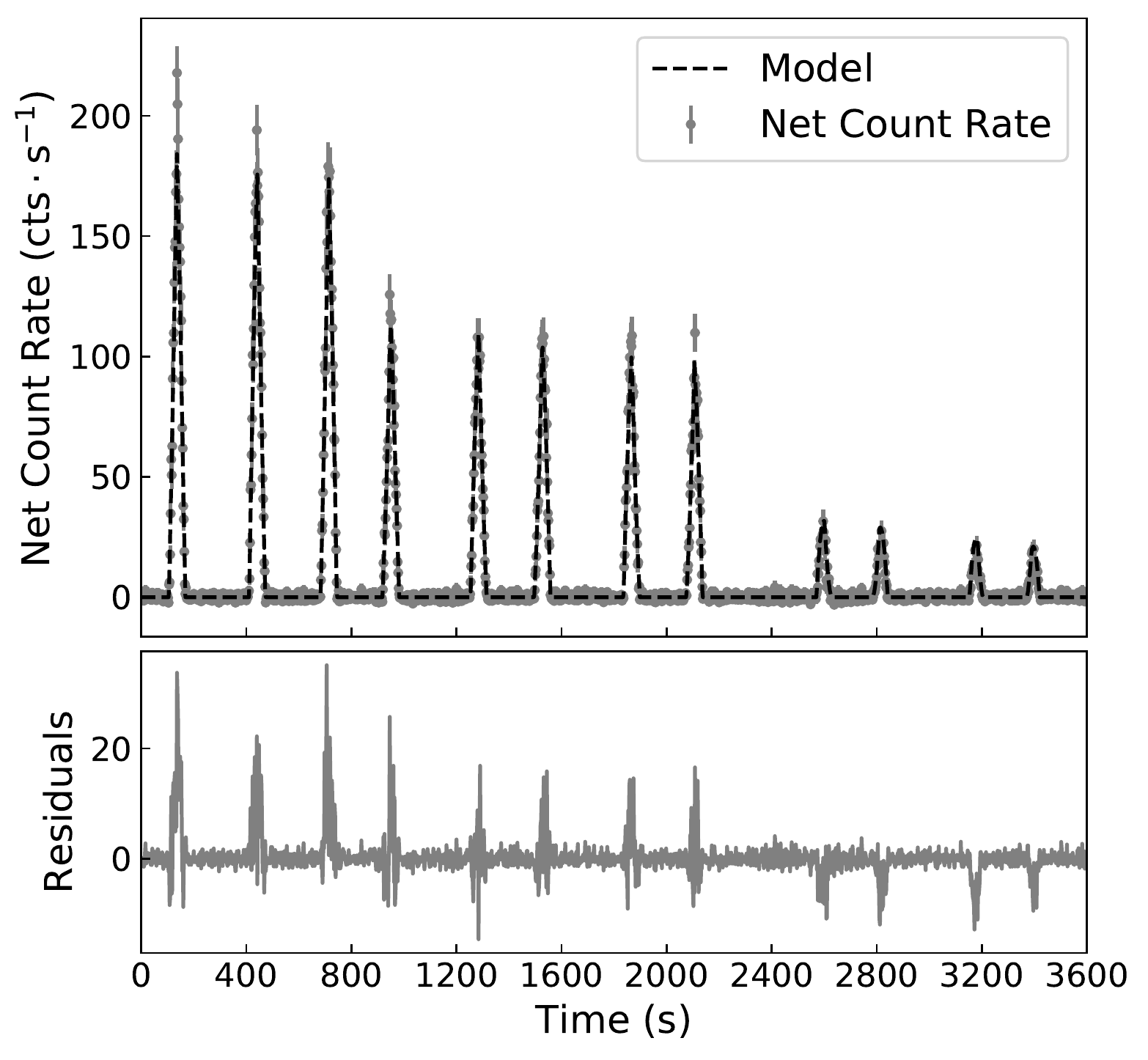}
    }
    \subfigure[LE: only with paraboloid correction]{
        \includegraphics[width=6 cm]{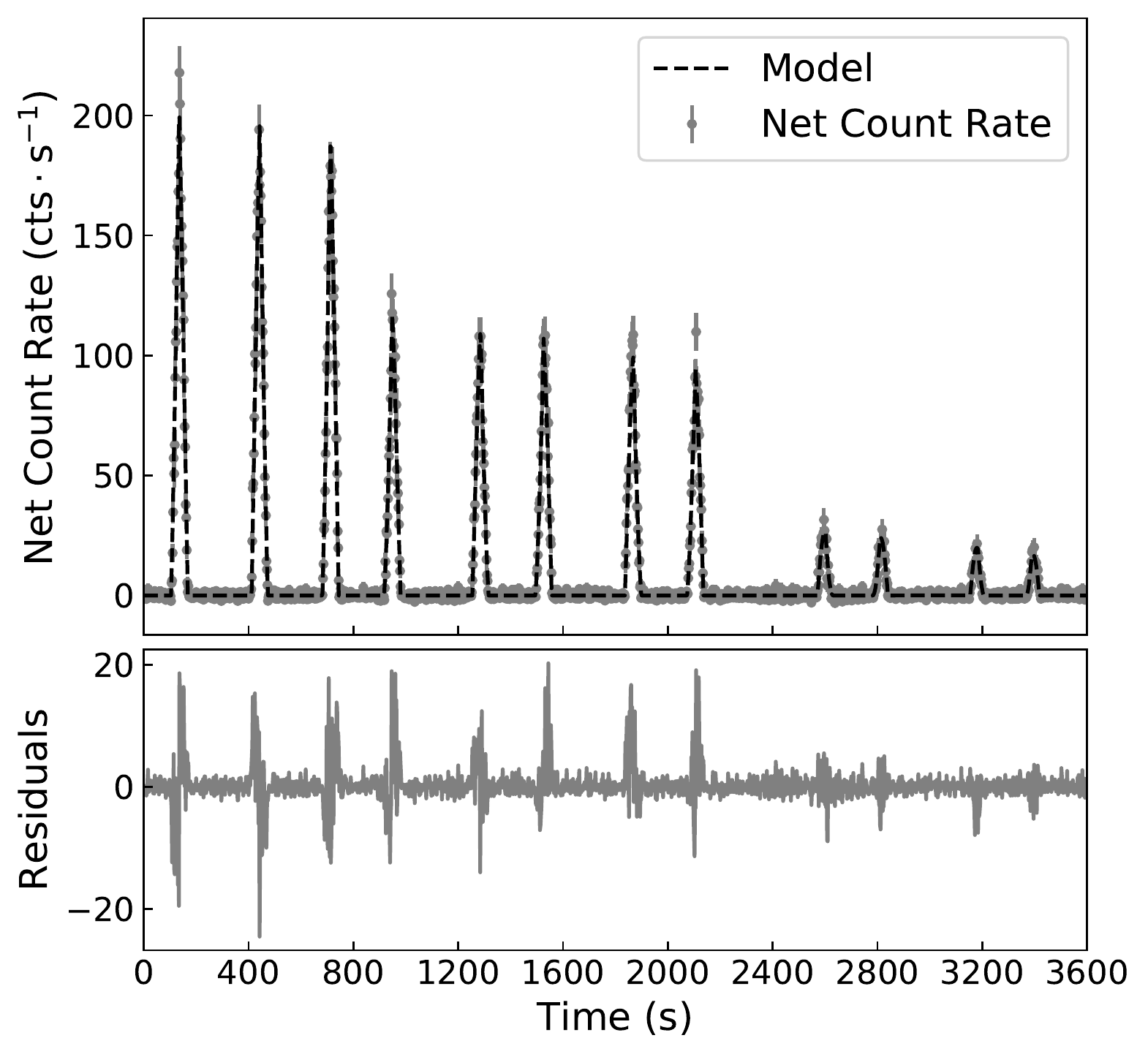}
    }
    \subfigure[ME: only with rotating correction]{
        \includegraphics[width=6 cm]{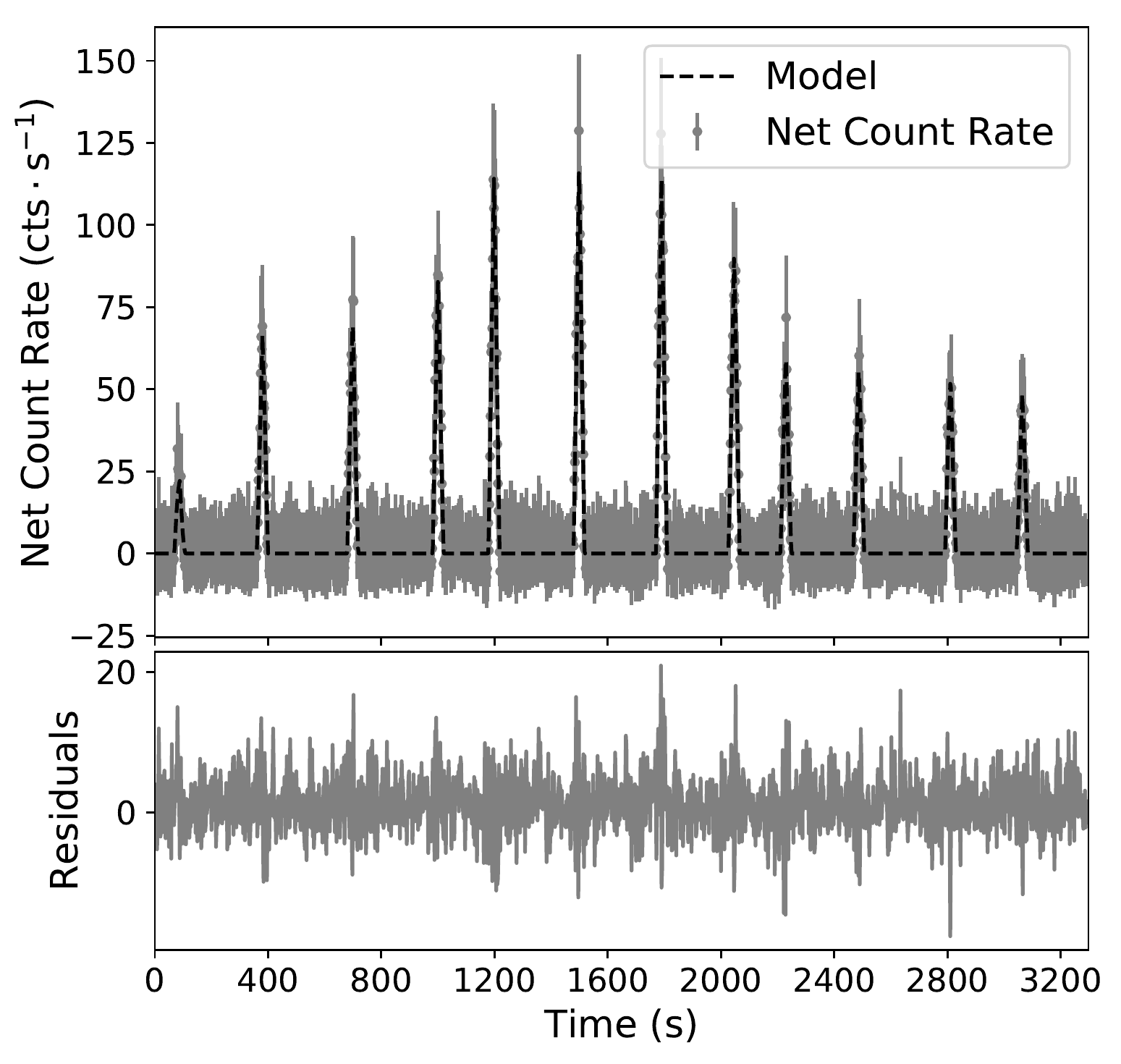}
    }
    \subfigure[ME: only with paraboloid correction]{
        \includegraphics[width=6 cm]{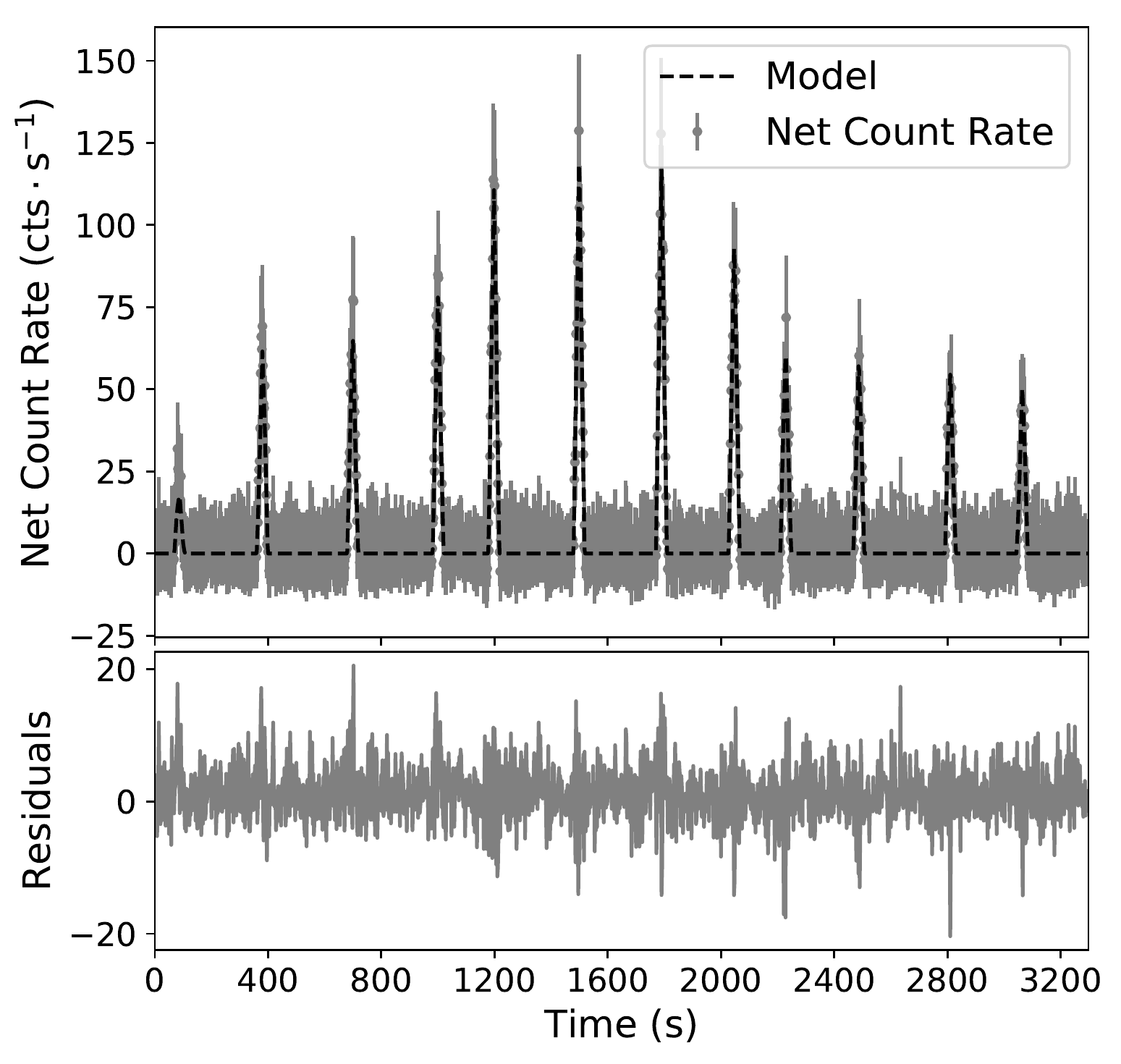}
    }
    \subfigure[HE: only with rotating correction]{
        \includegraphics[width=6 cm]{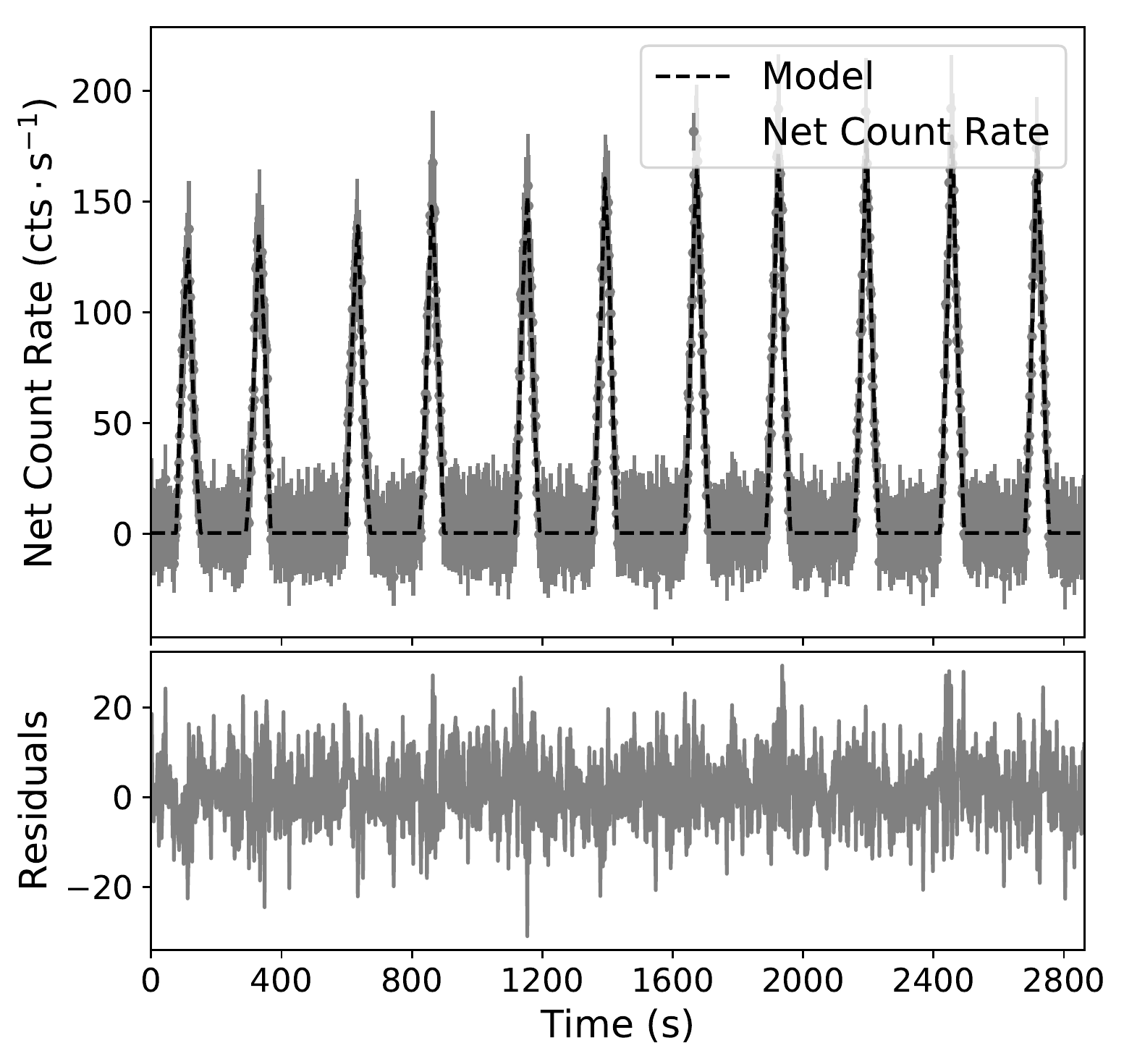}
    }
    \subfigure[HE: only with paraboloidal correction]{
        \includegraphics[width=6 cm]{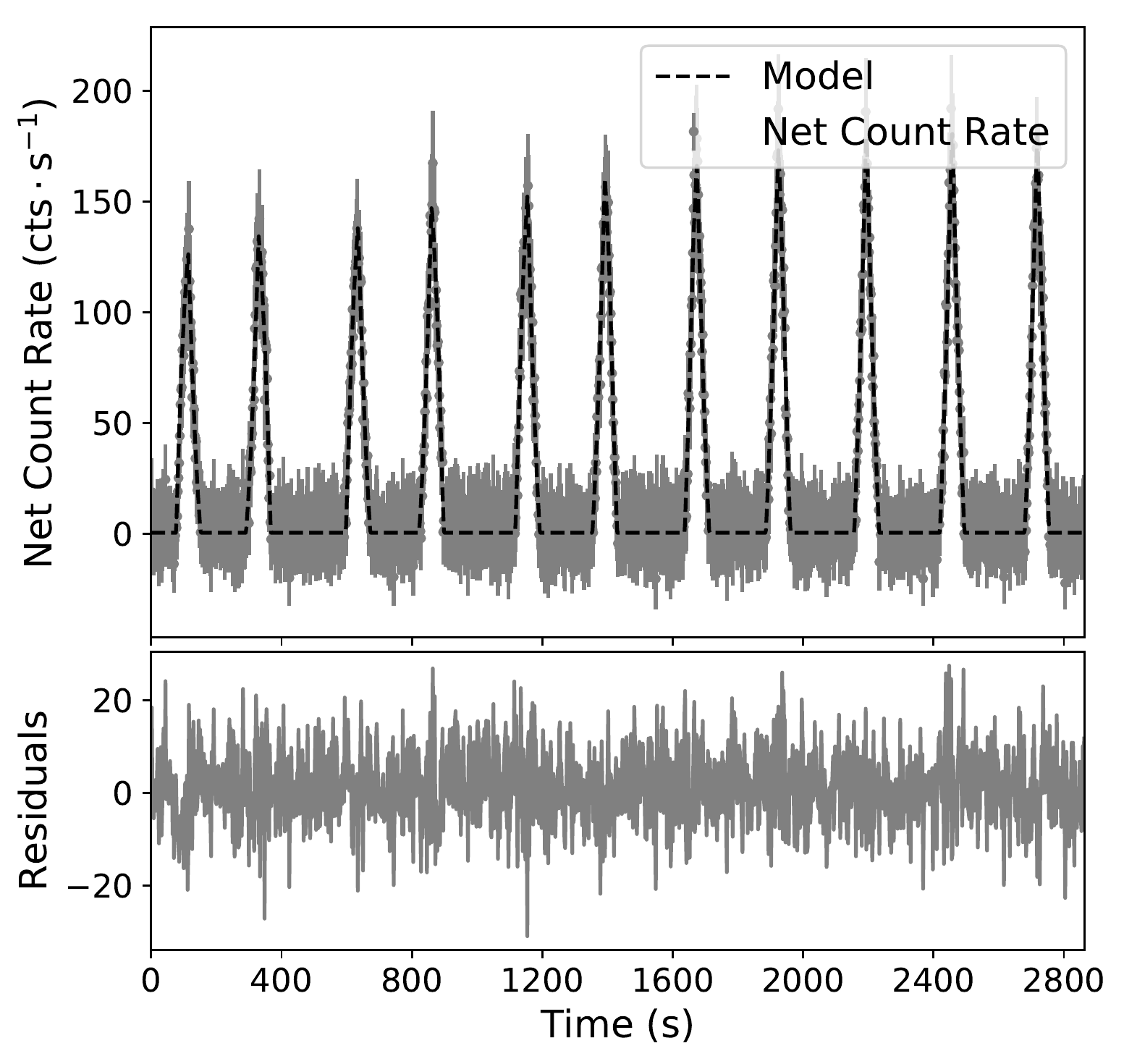}
    }
    \caption{Improvement of the PSFs after rotating correction and paraboloid correction separately. The data is the same as that in Figure \ref{fig_example}.}
    \label{fig_singlecorr}
\end{figure}

We fit the calibrating data with the corrected PSF described in Equation \ref{eq_psf_c} to determine the adjusting parameters (three Euler angles and four paraboloidal parameters). In the fittings, the position (R.A. and Dec.) of the Crab is fixed at its real position ($83^{\circ}.633,22^{\circ}.015$), and the normalized flux of the Crab is fixed at the mean count-rate measured by the \emph{Insight-HXMT} which are listed in Table \ref{comparison}. However the seven adjusting parameters are free parameters in the models in the fittings. Table \ref{tab_par} shows the seven parameters obtained from fitting. Fig. \ref{fig_fuc} shows the profiles of the correcting paraboloidal functions of the three telescopes. Fig. \ref{fig_new} shows the fitting result with the calibrated PSFs using the same data as Fig. \ref{fig_example}. In Fig. \ref{fig_new}, the goodness of fitting increases significantly with the corrected PSFs replacing the geometrical PSFs. Furthermore, the mirror symmetric feature and inhomogeneous feature existing in the residuals of Fig. \ref{fig_example} disappeared. Fig. \ref{fig_singlecorr} shows improvement of the PSFs after each correction; it can be seen that the mirror symmetric features can be improved by rotating correction and inhomogeneous features can be improved by paraboloidal correction in LE data.

\begin{table}[tp]
    \centering
    \caption{Parameters of the rotating matrix and paraboloidal function in PSF calibration}
    \begin{tabular}{ccccc}
        \toprule
        \multicolumn{2}{c}{Parameter} & LE& ME& HE\\
        \hline
        \multirow{3}{*}{Rotation}&$\psi (deg)$ &0.00&0.10&0.01\\
        &$\theta (deg)$ &0.12&-0.05&0.02\\
        &$\phi (deg)$ &0.09&0.17&0.04\\
        \midrule
        \multirow{4}{*}{Paraboloid}& $a$&-0.001&-0.038&0.000\\
        &$b$&-0.012&-0.001&-0.007\\
        &$c$&-0.043&0.218&0.077\\
        &$d$&1.023&0.918&0.964\\
        \bottomrule
        \end{tabular}
    \label{tab_par}
\end{table}

\begin{figure}
    \centering
    \includegraphics[width=15cm]{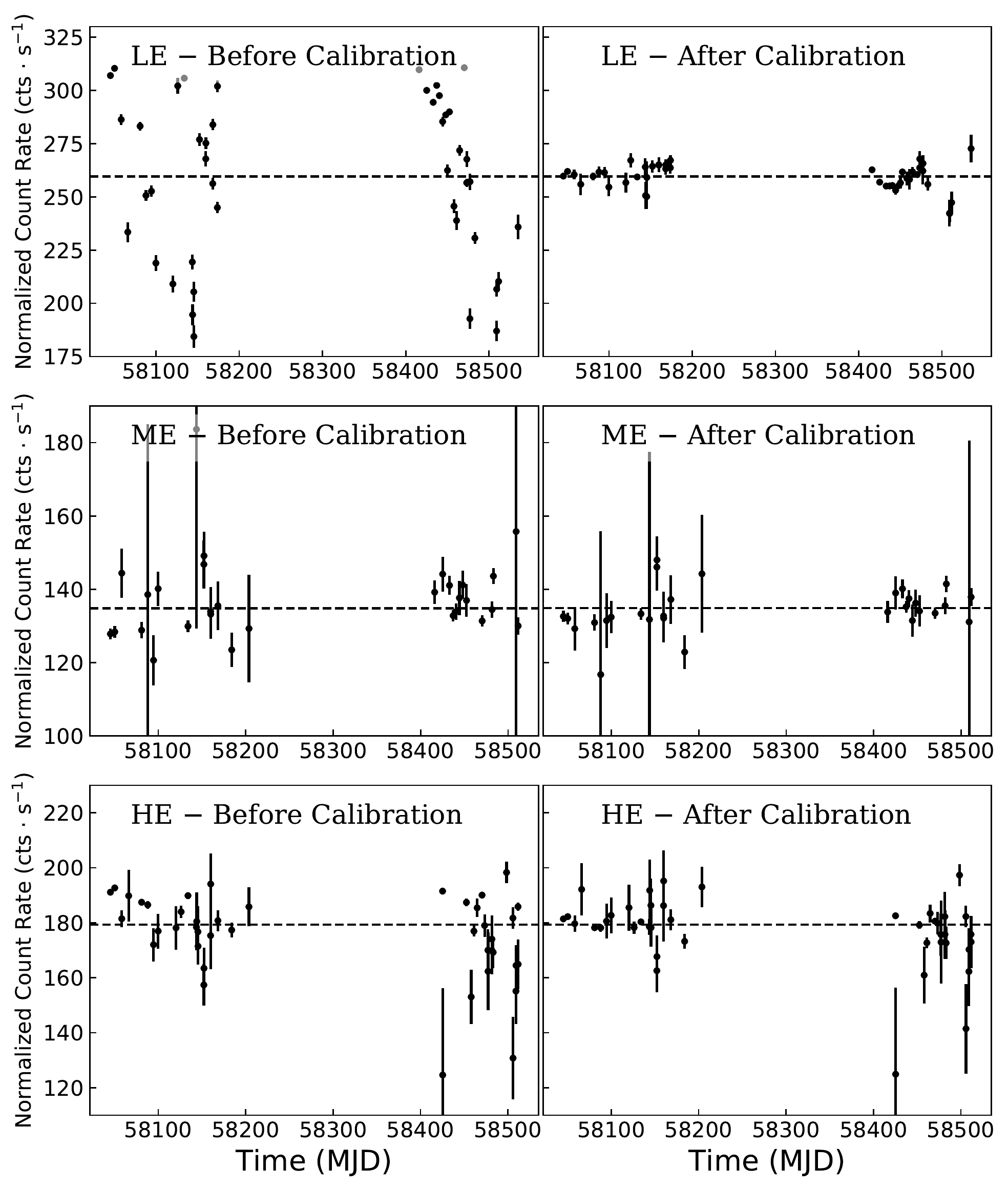}
    \caption{Comparison of the long-term light curve of the Crab obtained with geometrical PSF and calibrated PSF in the regular scanning observations. The dashed lines are the mean count rates of the Crab.}
    \label{fig_err_flx}
\end{figure}

\begin{figure}
    \centering
    \includegraphics[width=15cm]{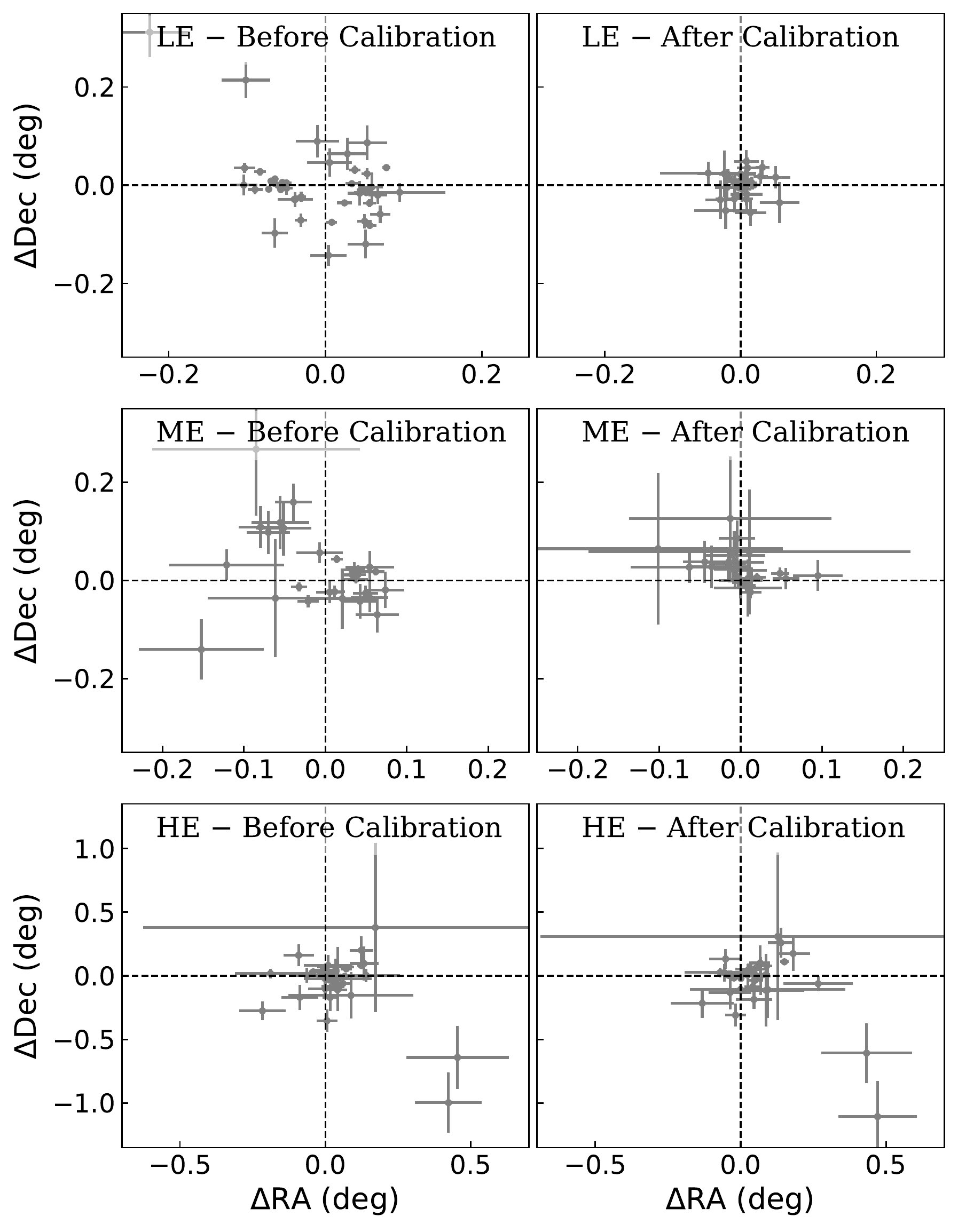}
    \caption{Comparison of the estimated positions of the Crab obtained with geometrical PSF and calibrated PSF in the regular scanning observations.}
    \label{fig_err_loc}
\end{figure}

The regular scanning observations of the Crab are used to test the calibrated PSFs. Figure \ref{fig_err_flx} shows the comparison of the long-term light curve of the Crab obtained with geometrical PSF (before calibration) and calibrated PSF (after calibration) in the regular Galactic plane scanning survey. The comparison of the estimated positions of the Crab obtained with the two PSFs are shown in Figure \ref{fig_err_loc}. It can be seen that the dispersions of both the position and normalized flux decrease significantly with the calibrated PSF. The systematic errors are the intrinsic dispersions of the data that can be calculated by solving the equation
\begin{gather}
    \sum_{i}^N \frac{(f_{i}- \bar{f})^{2}}{\sigma_{{\rm t},i}^{2}} = N-1,
    \label{eq_sys}
\end{gather}
where 
\begin{gather}
    \sigma_{{\rm t},i}^{2} = \sigma_{\rm sys}^{2} + \sigma_{{\rm stat},i}^{2},\\
    \bar{f} = \sum_{i}^N f_{i}\times w_{i}, \quad w_{i} = \frac{\frac{1}{\sigma_{{\rm t},i}^{2}}}{\sum_{i}^N \frac{1}{\sigma_{{\rm t},i}^{2}}},
\end{gather}
where $\sigma_{\rm sys}$ is the systematic error, $f_{i}$ the measured data, $\sigma_{{\rm stat},i}$ the statistic errors of data, and $\sigma_{{\rm t},i}$ the total errors of data. 

The systematic errors in the localization and flux estimation with the geometrical PSF and calibrated PSF are calculated. As shown in Table \ref{comparison}, the systematic errors in source localization decrease from $0^{\circ}.088$, $0^{\circ}.063$, $0^{\circ}.118$ to $0^{\circ}.010 $, $0^{\circ}.015$, $0^{\circ}.113$ for LE, ME and HE, and the systematic errors in flux estimation decrease from 14.8\%, 3.8\%, 5.7\% to 1.8\%, 1.6\%, 2.7\% for LE, ME and HE, respectively.

\begin{table*}[tp]
    \centering
    \caption{Comparison of the long-term observations of the Crab before and after the calibration}
    \label{comparison}
    \begin{threeparttable}
    \begin{tabular}{ccccccc}
    \toprule
	\multirow{2}{*}{Telescope}&\multicolumn{2}{c}{LE}&\multicolumn{2}{c}{ME}&\multicolumn{2}{c}{HE}\\
    \cmidrule(r){2-3} \cmidrule(r){4-5} \cmidrule(r){6-7}
    & before & after & before & after & before & after \\
    \midrule
         ${\Delta_{\rm m}{\rm RA}}$ (deg) & -0.013 &0.003 &0.009&0.008&0.018&0.029\\
         ${\Delta_{\rm m}{\rm Dec}}$ (deg)& -0.003 & 0.004 &0.013&0.002&-0.018&-0.009\\
         Mean Flux \tnote{*} $({\rm cts~s}^{-1})$& \multicolumn{2}{c}{259.6} &\multicolumn{2}{c}{134.9}&\multicolumn{2}{c}{179.3} \\
         $\sigma_{\rm sys,l} $\quad (deg)&0.088&0.010 &0.063&0.015&0.118&0.113\\
         $\sigma_{\rm sys,f} \quad ({\rm cts~s}^{-1})$&29.7&4.7 &4.3&2.2&10.2&4.9\\
        \bottomrule
        \end{tabular}
\begin{tablenotes}
\item Note: ${\Delta_{\rm m}{\rm RA}}$ denotes the mean value of the difference between the fitted R.A. and the real R.A. of the Crab in the tangent plane, ${\Delta_{\rm m}{\rm Dec}}$ denotes the mean value of the difference between the fitted Dec. and the real Dec. of the Crab in the tangent plane, $\sigma_{\rm sys,l}$ denotes the systematic error of the estimated position in the tangent plane, $\sigma_{\rm sys,f}$ denotes the systematic error of the estimated flux.
\item[*] The values are normalized to the count rates as the Crab is in the center of the FOVs.
\end{tablenotes}
\end{threeparttable}
\end{table*}

\section{Summary and Discussion}
We present empirical adjustments on the geometrical PSFs of LE , ME and HE of \emph{Insight-HXMT}. The calibration scanning observations of the Crab are used to calibrate the PSFs, and the regular scanning observations of the Crab are used to test the calibration. The adjusting method contains a rotating matrix and a paraboloidal function to multiply the theoretical calculated geometrical PSF. When considering variable backgrounds and signal to noise ratio in different energy ranges, we do not use the whole energy ranges of \emph{Insight-HXMT} in analyzing regular scanning observations in post-processing on ground, which has been noticed in Table \ref{tab_instr}. The inhomogeneous correcting parameters may vary with energies, while the rotating correction can be extended to the whole energy ranges.

For HE, the observed count rate of a scanned source is larger than that expected with the geometrical calculated PSF when the source is near the both edges of FOV in the short-side direction. This features can be attributed to the penetration of high energy photons. For other deformations of the PSFs, we can not identify the reasons precisely. Such as LE, we initially considered some mechanical adjustment into PSFs, like vacuum gaps between the detectors and tantalum grid of collimators, blank zones between the detectors and brackets of the detectors. It improve fittings to some extent. Nevertheless, the empirical adjustments work much better. The photons can be also reflected or scattered by the collimators wall of \emph{Insight-HXMT} to reach the detector that can increase the counts rate especially when a source near the edge of the FOV.

Although the exposure time of the calibrating scanning observations is more than 3~Ms, the calibrating data is still insufficient to construct a discrete PSF with highly dense micro-grids directly. Especially on the edges of FOVs, the count rate is low due to small detecting area, which results in large statistical errors than in the center of FOVs and affects the calibration. Therefore, some structures can be seen in fitting residuals with calibrated PSF in some observations. Moreover, the estimations to the position and flux of a source can deviate from the real values as a source is scanned cross the edge of a FOV. Hence, we screen out the results obtained from the observations where the source appears on an edge of a FOV. It is worth noting that this will not affect the pointing observations of \emph{Insight-HXMT}, because the target will not be away from the center of the FOV too far even in the off-axis observation.

For the Galactic plane scanning survey, after the calibration, the systematic errors in source localization are $0^{\circ}.010 $, $0^{\circ}.015$, $0^{\circ}.113$  for LE, ME and HE, and the systematic errors in flux estimation are 1.8\%, 1.6\%, 2.7\% for LE, ME and HE, respectively. Source flux in off-axis pointing observation also can be estimated accurately with the calibrated PSF. There is a possibility of the long-term evolution of the PSF, thus the PSF calibration will be performed continually in the future.

\section*{Acknowledgments}
This work made use of the data from the \emph{Insight-HXMT} mission, a project funded by China National Space Administration (CNSA) and the Chinese Academy of Sciences (CAS). The authors thank supports from the National Program on Key Research and Development Project (Grant No. 2016YFA0400802) and the National Natural Science Foundation of China under Grants No. U1838202, U1838201, 11703028, U1838105 and U1838113.


\begin{thebibliography}{99}
    \bibitem[Ackermann(2012)]{Ackermann2012}
    Ackermann, M., Ajello, M., Albert, A., et al., 2012. ApJS. 203, 4. DOI: 10.1088/0067-0049/203/1/4.
    \bibitem[Cao et al.(2019)]{Cxl2019}
    Cao, X.-L., Jiang, W.-C., Meng, B., et al., 2019. Sci. China-Phys. Mech. Astron. in press. arXiv:1910.04451.
    \bibitem[Chen et al.(2019)]{Cy2019}
    Chen, Y., Cui, W.-W., Li, W., et al., 2019. Sci. China-Phys. Mech. Astron. in press. arXiv:1910.08319. DOI: 10.1007/s11433-019-1469-5.
    \bibitem[Hiroi et al.(2013)]{Hiroi2013}
    Hiroi, K., Ueda, Y., Hayashida, M., et al., 2013. ApJS. 207, 36. DOI: 10.1088/0067-0049/207/2/36.
    \bibitem[Li et al.(2009)]{Li2009}
    Li, G., Wu, M., Zhang, S., Jin, Y.-K., 2009. Chin. Astron. Astrophys. 33, 333. DOI: 10.1016/j.chinastron.2009.07.013.
    \bibitem[Li and Wu(1993)]{Li1993}
    Li, T.-P., Wu, M., 1993. Ap\&SS. 206, 91. DOI: 10.1007/BF00658385.
    \bibitem[Li and Wu(1994)]{Li1994}
    Li, T.-P., Wu, M., 1994. Ap\&SS. 215, 213. DOI: 10.1007/BF00660079.
    \bibitem[Liu et al.(2019)]{Lcz2019}
    Liu, C.-Z., Zhang, Y.-F., Li, X.-F., et al., 2019. Sci. China-Phys. Mech. Astron. in press. arXiv:1910.04955.
    \bibitem[Madsen et al.(2015)]{Madsen2015}
    Madsen, K.K., Harrison, F.A., Markwardt, C.B., et al., 2015. ApJS. 220, 8. DOI: 10.1088/0067-0049/220/1/8.
    \bibitem[Morh{\'{a}}{\v{c}} et al.(1997)]{Morh1997}
    Morh{\'{a}}{\v{c}}, M., Kliman, J., Matou{\v{s}}ek, V., et al., 1997. Nucl. Instrum. Method. Phys. Res. Sect. A. 401, 113. DOI: 10.1016/S0168-9002(97)01023-1.
    \bibitem[Roy et al.(1977)]{Roy1977}
    Roy, A., Ballas, J., Jagoda, N., et al., 1977. IEEE Transactions on Nuclear Science 24, 804. DOI: 10.1109/TNS.1977.4328786.
    \bibitem[Ryan et al.(1988)]{Ryan1988}
    Ryan, C., Clayton, E., Griffin, W., et al., 1988. Nucl. Instrum. Method. Phys. Res. Sect. B. 34, 396. DOI: 10.1016/0168-583X(88)90063-8.
    \bibitem[Xie et al.(2015)]{Xie2015}
    Xie, F., Zhang, J., Song, L.-M., et al., 2015. Ap\&SS. 360, 47. DOI: 10.1007/s10509-015-2559-1.
    \bibitem[Zhang et al.(2019)]{Zsn2019}
    Zhang, S.-N., Li, T.-P., Lu, F.-J., et al., 2019. Sci. China-Phys. Mech. Astron. in press. arXiv:1910.09613. DOI: 10.1007/s11433-019-1432-6.

\end{thebibliography}
\end{document}